\documentclass[11pt,a4paper]{article}
\pdfoutput=1
\usepackage{ifpdf}
\usepackage{jheppub}
\usepackage{rotating}
\usepackage[bb=boondox]{mathalfa}
\usepackage{comment}

\usepackage{tikz-cd}

\usepackage{graphicx} % Required for inserting images
\usepackage{mathrsfs}
\usepackage{amsmath,amsfonts,amssymb}
\usepackage{mathtools}
 \usepackage{booktabs}
\usepackage{graphicx}% Include figure files
\usepackage{dcolumn}% Align table columns on the decimal point
\usepackage{bm}% bold math
\usepackage{hyperref}% add hypertext capabilities
\usepackage[mathlines]{lineno}% Enable numbering of text and display math
\usepackage{float}
\usepackage{blindtext}
\usepackage{titlesec}
\title{Sections and Chapters}
\usepackage[toc,page]{appendix}
\usepackage{xcolor}
\usepackage{graphicx}
\graphicspath{ {./images/} }
\usepackage[thinc]{esdiff}
\usepackage[euler]{textgreek}
\usepackage{braket}
\usepackage{physics}
\usepackage{xfrac}
\usepackage{soul}
\usepackage{stmaryrd}
\usepackage{pifont}
\usepackage{caption}
\usepackage{makecell}
\usepackage{orcidlink}
\usepackage{soul}
\usepackage{multirow}

\allowdisplaybreaks

\title{\textcolor{black}{Two-Component Dark Matter and Strongly First-Order Electroweak Phase Transition via a Two-Step Mechanism in the Inert Doublet plus Real Singlet Model}}

\author[a]{Deeksha Dattaram Naik\,\orcidlink{0000-0000-0000-0000}\,}
\author[a]{\!\!, Arnab Chaudhuri\,\orcidlink{0009-0009-9592-4786}\,}
\affiliation[a]{School of Advanced Sciences,
Vellore Institute of Technology, Vellore, Tamil Nadu 632014, India.}
\emailAdd{deekshadattaram.naik@vit.ac.in}
\emailAdd{arnab.chaudhuri@vit.ac.in}

\abstract{We study a two-component dark matter model in which a real singlet scalar and an inert doublet scalar are simultaneously present alongside the Standard Model, each stabilised by an independent discrete symmetry. The real singlet undergoes a strong first-order transition at high temperature, establishing an intermediate phase with a non-zero singlet vacuum expectation value, which in turn generates the barrier required for the Higgs field to complete a strongly first-order electroweak transition at a lower temperature. This two-step mechanism naturally enforces the condition $v_c/T_c \geq 1$, the sphaleron washout-avoidance criterion necessary for electroweak baryogenesis. The gravitational wave spectrum from both steps is computed; the second step produces a decihertz signal, peaking near $0.08$--$0.34~\mathrm{Hz}$ across the viable parameter space and $0.14$--$0.22~\mathrm{Hz}$ at the benchmark points, that reaches the sensitivity of DECIGO for the strongest transitions and of BBO more widely, while the first remains sub-threshold for all planned detectors. At the coupling values required to drive the phase transition, the singlet is generically underabundant as a thermal relic, so the inert doublet carries the dominant fraction of the observed relic density. Confronting the model with current constraints, we find that the LUX-ZEPLIN 2024 result excludes the entire electroweak-phase-transition-viable parameter space of the singlet sector across the full mass range, even after rescaling by the singlet relic fraction, establishing a complete no-go for the singlet as a viable thermal dark matter candidate in this framework.}

\begin{document}
\maketitle
\flushbottom

\preprint{}

\section{Introduction}
\label{sec:intro}

The origin of the baryon asymmetry remains one of the central open questions 
in particle physics and cosmology. Electroweak 
baryogenesis~\cite{Kuzmin:1985mm, Arnold:1987mh, Shaposhnikov:1987tw} is 
particularly attractive because it operates entirely at the electroweak 
scale and is therefore testable with current and near-future experiments. It 
requires a strongly first-order transition satisfying $v_c/T_c \geq 1$, so 
that sphaleron processes are suppressed inside the nucleated bubbles and the 
generated asymmetry is not washed out~\cite{Rubakov:1996vz, Morrissey:2012db, 
White:2016nbo}. This condition is necessary but not sufficient: a complete 
realisation also demands a CP-violating source at the bubble wall, which the 
real scalar couplings of the model studied here do not supply, and we 
therefore treat $v_c/T_c \geq 1$ throughout as the sphaleron 
washout-avoidance condition without computing a baryon asymmetry. Since the 
Standard Model alone yields a smooth crossover at 
$m_h = 125~\mathrm{GeV}$~\cite{Kajantie:1996mn, Csikor:1998eu, Aoki:1999fi,Chaudhuri:2017icn}, 
new scalar degrees of freedom are required.

The real singlet extension is the most economical 
option~\cite{McDonald:1993ey, Espinosa:1993bs, Espinosa:1996qw, 
Espinosa:2007qk, Profumo:2007wc, Espinosa:2011ax, Carena:2019une, 
Ellis:2022lft,Chaudhuri:2022sis}: the singlet couples to the Higgs through a renormalisable 
portal, supplies a dark matter candidate when protected by a discrete 
symmetry~\cite{McDonald:1993ex, Burgess:2000yq, Davoudiasl:2004be, 
Patt:2006fw}, and can generate an observable gravitational wave 
signal~\cite{Caprini:2019egz, Espinosa:2010hh, Hindmarsh:2015qta, 
Hindmarsh:2016lnk, Weir:2017wfa,Chaudhuri:2024vrd,Srivastava:2025oer,Chaudhuri:2025ybh,Chaudhuri:2026ovn,Chaudhuri:2025cjp}. It also carries a well-known tension: the 
portal coupling needed for a strong transition is generically of order one, 
whereas direct detection bounds it to $\lambda_{H\phi} \lesssim 
10^{-3}$--$10^{-4}$ over most of the mass range~\cite{XENON:2018voc, 
LUX:2016ggv, Aalbers:2022dzr}, a constraint sharpened considerably by 
LUX-ZEPLIN 2024~\cite{LZ:2024zvo}. The inert doublet 
model~\cite{Deshpande:1977rw, Barbieri:2006dq, Ma:2006km} offers a 
complementary dark sector whose lightest neutral component is 
stable~\cite{LopezHonorez:2006gr, Dolle:2009fn, LopezHonorez:2010eeh, 
Arhrib:2013ela, Belyaev:2016lok}, which modifies the transition through loop 
contributions and supports multi-step histories of its 
own~\cite{Blinov:2015vma, Fabian:2020hny, Benincasa:2022elt}, subject to 
direct detection, Higgs invisible width~\cite{ATLAS:2023tkt}, and 
electroweak precision constraints~\cite{Pierce:2007ut}. Two-component 
constructions combining such sectors have been explored 
widely~\cite{Belanger:2014bga, Bhattacharya:2016ysw, DuttaBanik:2018emv, 
Bhattacharya:2019fgs, Liu:2017gfg, Borah:2024emz,Basak:2021tnj}, as has the interplay 
between transition dynamics and dark matter 
phenomenology~\cite{Huang:2017rzf, Bell:2020gug, Zhang:2023mnu, Srivastava:2025oer}.

Combining these two sectors turns out to be more than the sum 
of its parts, because the combination makes a tension unavoidable. The 
two-step mechanism does not merely permit a large portal coupling; it 
requires one, $\lambda_{H\phi} \gtrsim 0.32$ throughout the region that 
supports a genuine two-step thermal history. That same coupling fixes the 
singlet nucleon cross section. The central result of this work is that the 
two demands are irreconcilable: at every parameter point completing the sequence, the relic-fraction-rescaled
cross section exceeds the LZ-2024 bound, by four to five orders of magnitude
away from the Higgs resonance and by a narrower margin at
$m_\phi \simeq m_h/2$, where resonant annihilation depletes the singlet
abundance. The light window $m_\phi < m_h/2$ is closed independently by the
Higgs invisible width. The 
singlet cannot both trigger the strongly first-order transition and survive 
as a thermal relic. This is a structural exclusion rather than a statement 
about a corner of parameter space, and it is what forces the two-component 
structure: the division of labour between the sectors is an output of the 
constraints, not an assumption of the model building.

Two consequences follow. First, the experimental roles invert 
relative to the usual Higgs-portal narrative. Direct detection, ordinarily 
the discovery channel, functions here purely as the exclusion arm and has 
already completed its work, while leaving the transition mechanism itself 
untouched, since that mechanism never required the singlet to be stable 
relic matter. Discovery capability migrates entirely to the decihertz 
gravitational wave band. Second, that prediction is quantitatively fragile 
in a way not consistently appreciated. Two-step transitions complete 
rapidly, with $\beta/H \sim 6\times10^3$--$8\times10^3$ here, and in this 
regime the finite lifetime of the acoustic source suppresses the sound-wave 
amplitude by two to three orders of magnitude through 
$\Upsilon(\tau_{\rm sw})$. Because $\Upsilon$ itself scales as 
$(\beta/H)^{-1}$, the peak amplitude falls as $(\beta/H)^{-2}$ rather than 
$(\beta/H)^{-1}$, so detectability claims for fast transitions that omit 
this factor overstate the signal by orders of magnitude. We include it 
throughout, and our statements are correspondingly conservative.

The mechanism producing these results operates as follows. The real singlet, 
protected by $\mathbb{Z}_2^{(1)}$, undergoes a first-order transition at a 
high temperature $T_1$, acquiring a vacuum expectation value and 
establishing an intermediate phase in which electroweak symmetry remains 
unbroken. This phase reshapes the Higgs effective potential so that when the 
singlet returns to the origin at a lower temperature $T_2$, the electroweak 
transition is strongly first-order. We emphasise that 
$\xi_H \geq 1$, where $\xi_H$ is $v_c/T_c$, is an output rather than an input: every point completing the 
two-step sequence satisfies it without that condition being imposed as a 
selection criterion. The inert doublet, protected by an independent 
$\mathbb{Z}_2^{(2)}$, does not participate in the transition but supplies 
the dark matter, its abundance set by gauge interactions and 
therefore decoupled from the portal coupling that direct detection 
constrains.

We compute the gravitational wave spectrum from both steps using the sound 
wave, turbulence, and collision contributions~\cite{Espinosa:2010hh, 
Caprini:2019egz, Hindmarsh:2015qta, Hindmarsh:2016lnk, Cutting:2019zws}, 
comparing against DECIGO~\cite{Seto:2001qf, Kawamura:2011zz}, LISA~\cite{LISA:2017pwj, Caprini:2015zlo}, 
BBO~\cite{Harry:2006fi}, and ET~\cite{Punturo:2010zz}. The second step 
peaks near $0.08$--$0.34~\mathrm{Hz}$ across the scanned parameter space, 
rising above the DECIGO sensitivity for the strongest transitions and 
reached by BBO more widely, while the first step remains sub-threshold 
everywhere. The model is then confronted with 
LZ-2024~\cite{LZ:2024zvo}, the ATLAS invisible width 
combination~\cite{ATLAS:2023tkt}, electroweak precision 
observables~\cite{ParticleDataGroup:2022pth}, and LEP2 charged scalar 
searches~\cite{Lundstrom:2008ai}, establishing the no-go described above 
and complementing earlier analyses of the direct detection versus EWPT 
tension~\cite{Carena:2019une, Beniwal:2017eik, Falkowski:2012fb}.

The paper is organised as follows. Section~\ref{sec:model} introduces the 
model and mass spectrum, Section~\ref{sec:ewpt} the finite-temperature 
effective potential and two-step mechanism, Section~\ref{sec:gw} the 
gravitational wave spectrum, and Section~\ref{sec:dm} the two-component 
dark matter phenomenology. Section~\ref{sec:conclusions} summarises. 
Appendix~\ref{app:veff} collects the effective potential expressions and 
Appendix~\ref{app:paramscan} the parameter scan details.

\section{The Model}
\label{sec:model}

\subsection{Field content and discrete symmetries}

We extend the Standard Model scalar sector by two additional fields: a real 
gauge singlet $\phi$ and an inert $\mathrm{SU}(2)_L$ doublet $\eta$ with 
hypercharge $Y = 1/2$. The stability of both fields as dark matter candidates 
is ensured by imposing two independent $\mathbb{Z}_2$ symmetries. Under 
$\mathbb{Z}_2^{(1)}$, the singlet transforms as $\phi \to -\phi$ while all 
other fields are even. Under $\mathbb{Z}_2^{(2)}$, the inert doublet 
transforms as $\eta \to -\eta$ while all other fields are even. The two 
symmetries are independent and unbroken at zero temperature, so neither $\phi$ 
nor $\eta$ acquires a vacuum expectation value at $T = 0$, and both are stable 
on cosmological timescales. The field content and charge assignments are 
summarised in Table~\ref{tab:fields}.

\begin{table}[h]
\centering
\begin{tabular}{ccccc}
\hline\hline
Field & $\mathrm{SU}(2)_L$ & $U(1)_Y$ & $\mathbb{Z}_2^{(1)}$ & 
$\mathbb{Z}_2^{(2)}$ \\
\hline
$H$   & $\mathbf{2}$ & $+1/2$ & $+$ & $+$ \\
$\phi$ & $\mathbf{1}$ & $0$    & $-$ & $+$ \\
$\eta$ & $\mathbf{2}$ & $+1/2$ & $+$ & $-$ \\
\hline\hline
\end{tabular}
\caption{Scalar field content and discrete charge assignments. 
Standard Model fields carry charge $+$ under both symmetries.}
\label{tab:fields}
\end{table}

\subsection{The scalar potential}

The most general renormalisable scalar potential consistent with the gauge 
symmetry and both $\mathbb{Z}_2$ symmetries is
\begin{equation}
V = V_H + V_\phi + V_\eta + V_{\rm mix},
\label{eq:potential}
\end{equation}
where
\begin{align}
V_H &= -\mu_H^2\,|H|^2 + \lambda_H\,|H|^4, \\[4pt]
V_\phi &= \frac{1}{2}\mu_\phi^2\,\phi^2 
+ \frac{1}{4}\lambda_\phi\,\phi^4, \\[4pt]
V_\eta &= \mu_\eta^2\,|\eta|^2 
+ \lambda_\eta\,|\eta|^4 
+ \lambda_3\,|H|^2|\eta|^2 
+ \lambda_4\,|H^\dagger\eta|^2
+ \frac{\lambda_5}{2}\left[(H^\dagger\eta)^2 + \mathrm{h.c.}\right],
\label{eq:Veta}
\end{align}
and
\begin{equation}
V_{\rm mix} = \frac{\lambda_{H\phi}}{2}\,\phi^2\,|H|^2
+ \frac{\lambda_{\phi\eta}}{4}\,\phi^2\,|\eta|^2.
\label{eq:Vmix}
\end{equation}
With $\lambda_{H\phi} > 0$ and $\lambda_{\phi\eta} > 0$ 
at all benchmark points, the singlet direction adds only positive 
contributions to the quartic form, so the pairwise copositivity 
conditions above are also sufficient to guarantee that the full 
three-field potential is bounded below in every direction of field 
space. We follow the standard IDM 
convention~\cite{Deshpande:1977rw, Barbieri:2006dq, LopezHonorez:2006gr}; 
$\lambda_5$ is taken real without loss of generality, since its phase can 
be absorbed by a rephasing of $\eta$, and $\lambda_4$ is real by 
hermiticity. The effective portal coupling
\begin{equation}
\lambda_{H\eta}^{\rm eff} \equiv \lambda_3 + \lambda_4 + \lambda_5
\label{eq:leff}
\end{equation}
controls the $\eta^0_R$--Higgs interaction and governs both the 
Higgs-portal annihilation cross section and the direct detection cross 
section of $\eta^0_R$, as discussed in Section~\ref{sec:dm}.

The $\mathbb{Z}_2^{(1)}$ symmetry forbids all odd powers of $\phi$ and 
mixed terms linear in $\phi$ such as $\phi|H|^2$ or $\phi|\eta|^2$. The 
$\mathbb{Z}_2^{(2)}$ symmetry forbids odd powers of $\eta$ and any 
$H$--$\eta$ mixing term such as $H^\dagger\eta$. Together, the two 
symmetries eliminate all renormalisable mixing between the singlet and 
inert doublet sectors, except through the quartic portal 
$\phi^2|\eta|^2$ in Eq.~\eqref{eq:Vmix}. This coupling, parametrised by 
$\lambda_{\phi\eta}$, allows energy exchange between the two dark sectors 
during the thermal history and is discussed further in 
Section~\ref{subsec:relic}. The sign of $\mu_\phi^2$ is left free; the 
condition $\mu_\phi^2 < 0$ is required for the singlet to develop a 
non-zero expectation value at high temperature and trigger Step~I of the 
phase transition, as detailed in Section~\ref{sec:ewpt}.

\subsection{Physical mass spectrum}

Working in the unitary gauge, the SM Higgs doublet and the inert doublet 
are decomposed as
\begin{equation}
H = \begin{pmatrix} G^+ \\ \dfrac{v + h + iG^0}{\sqrt{2}} \end{pmatrix},
\qquad
\eta = \begin{pmatrix} \eta^+ \\ \dfrac{\eta^0_R + i\eta^0_I}{\sqrt{2}} 
\end{pmatrix},
\label{eq:unitarygauge}
\end{equation}
where $v = 246~\mathrm{GeV}$ is the Higgs vacuum expectation value, $h$ is 
the physical Higgs boson, and $G^\pm$, $G^0$ are the Goldstone bosons. The 
fields $\eta^0_R$, $\eta^0_I$, and $\eta^\pm$ are the physical inert scalars; 
they do not mix with the SM sector because $\mathbb{Z}_2^{(2)}$ forbids any 
$H$--$\eta$ mixing.

The Higgs potential minimum condition gives $\mu_H^2 = \lambda_H v^2$, from 
which the Higgs boson mass is
\begin{equation}
m_h^2 = 2\lambda_H v^2,
\qquad
\lambda_H = \frac{m_h^2}{2v^2} \simeq 0.129.
\label{eq:mh}
\end{equation}

The singlet mass at zero temperature follows from the second derivative of 
the potential at the electroweak vacuum $(\langle h\rangle = v,\, 
\langle\phi\rangle = 0)$:
\begin{equation}
m_\phi^2 = \mu_\phi^2 + \frac{\lambda_{H\phi}}{2}\,v^2.
\label{eq:mphi}
\end{equation}
Since we require $m_\phi^2 > 0$ at zero temperature but $\mu_\phi^2 < 0$ 
for Step~I of the phase transition, Eq.~\eqref{eq:mphi} gives the 
necessary condition on the portal coupling:
\begin{equation}
\lambda_{H\phi} > \frac{2m_\phi^2}{v^2}.
\label{eq:lHphi_condition}
\end{equation}
In practice this is automatically satisfied for our benchmark values.

The four IDM scalar masses are\,\footnote{%
These follow directly from expanding Eq.~\eqref{eq:Veta} at 
$\langle H\rangle = v/\sqrt{2}$. The charged mass receives only the 
$\lambda_3$ contribution because the $(H^\dagger\eta)^2$ and 
$|H^\dagger\eta|^2$ terms vanish for a charged component in the unitary 
gauge. The neutral masses split according to the relative sign of 
$\lambda_5$.}
\begin{align}
m_{\eta^0_R}^2 &= \mu_\eta^2 + \frac{\lambda_{H\eta}^{\rm eff}}{2}\,v^2,
\label{eq:meta0R} \\
m_{\eta^0_I}^2 &= \mu_\eta^2 + \frac{\lambda_3 + \lambda_4 - \lambda_5}{2}\,v^2,
\label{eq:meta0I} \\
m_{\eta^\pm}^2 &= \mu_\eta^2 + \frac{\lambda_3}{2}\,v^2,
\label{eq:metapm}
\end{align}
where $\lambda_{H\eta}^{\rm eff}$ is defined in Eq.~\eqref{eq:leff}. 
The condition $\lambda_5 < 0$ ensures $m_{\eta^0_R} < m_{\eta^0_I}$, 
so $\eta^0_R$ is the lightest $\mathbb{Z}_2^{(2)}$-odd state and the 
dark matter candidate. The mass splittings are
\begin{align}
m_{\eta^\pm}^2 - m_{\eta^0_R}^2 
&= -\frac{\lambda_4 + \lambda_5}{2}\,v^2,
\label{eq:Deltam} \\
m_{\eta^0_I}^2 - m_{\eta^0_R}^2 
&= -\lambda_5\,v^2.
\label{eq:deltam}
\end{align}
For $|\lambda_4|, |\lambda_5| \ll \lambda_3$, both splittings are small 
compared to the overall IDM mass scale, producing a compressed spectrum. 
Inverting Eqs.~\eqref{eq:meta0R}--\eqref{eq:metapm}, the soft mass 
parameter is determined by the lightest neutral mass:
\begin{equation}
\mu_\eta^2 = m_{\eta^0_R}^2 - \frac{\lambda_{H\eta}^{\rm eff}}{2}\,v^2.
\label{eq:mueta}
\end{equation}

The couplings of the physical Higgs boson $h$ to the BSM scalars are 
derived by substituting Eq.~\eqref{eq:unitarygauge} into the potential 
and collecting the trilinear terms. The relevant vertices are\,\footnote{%
The vertex $g_{hXX}$ is defined as the coefficient of $h\cdot X \cdot X$ 
in the interaction Lagrangian, so that the Feynman rule is $-ig_{hXX}$.}
\begin{align}
g_{h\phi\phi} &= \lambda_{H\phi}\,v, 
\label{eq:ghphiphi} \\
g_{h\eta^0_R\eta^0_R} &= \lambda_{H\eta}^{\rm eff}\,v, 
\label{eq:gheta0R} \\
g_{h\eta^0_I\eta^0_I} &= (\lambda_3 + \lambda_4 - \lambda_5)\,v, 
\label{eq:gheta0I} \\
g_{h\eta^+\eta^-} &= \lambda_3\,v.
\label{eq:ghetapm}
\end{align}
These couplings enter the Higgs invisible width (when $m_\phi < m_h/2$ 
or $m_{\eta^0_R} < m_h/2$), the one-loop Higgs self-energy, and the 
Higgs-portal dark matter annihilation and direct detection cross sections 
computed in Section~\ref{sec:dm}. The inter-sector quartic vertex 
$\phi^2\eta^0_R{}^2$ from $V_{\rm mix}$ carries a coupling strength 
$\lambda_{\phi\eta}/2$ and opens the conversion channel 
$\phi\phi \leftrightarrow \eta^0_R\eta^0_R$; its effect on the coupled 
relic density calculation is discussed in Section~\ref{subsec:relic}.

\subsection{Theoretical and electroweak precision constraints}

The potential in Eq.~\eqref{eq:potential} must satisfy vacuum stability, 
perturbativity, and perturbative unitarity. Stability in all field directions 
requires~\cite{Kannike:2012pe}
\begin{align}
\lambda_H > 0, \quad \lambda_\phi > 0, \quad \lambda_\eta > 0, \\
\lambda_{H\phi} > -2\sqrt{\lambda_H\lambda_\phi}, \\
\lambda_3 > -2\sqrt{\lambda_H\lambda_\eta}, \\
\lambda_3 + \lambda_4 - |\lambda_5| > -2\sqrt{\lambda_H\lambda_\eta}.
\label{eq:stability}
\end{align}
Perturbativity requires all dimensionless couplings to satisfy 
$|\lambda_i| < 4\pi$, and perturbative unitarity of $2 \to 2$ scalar 
scattering requires the eigenvalues of the scattering matrix to be bounded 
by $8\pi$~\cite{Lee:1977eg, Cynolter:2004cq}. All benchmark parameter 
sets in this paper satisfy these conditions.

At the level of electroweak precision observables, the real singlet $\phi$ 
does not contribute to the oblique parameters $S$, $T$, $U$ at tree level, 
since it is a gauge singlet. The IDM contributes to the $T$ parameter 
at one loop through
\begin{equation}
\Delta T = \frac{1}{16\pi^2\alpha_{\rm em} v^2}
\Big[F(m_{\eta^\pm}^2,\,m_{\eta^0_I}^2)
+ F(m_{\eta^\pm}^2,\,m_{\eta^0_R}^2)
- F(m_{\eta^0_I}^2,\,m_{\eta^0_R}^2)\Big],
\label{eq:DeltaT}
\end{equation}
where
\begin{equation}
F(m_1^2, m_2^2) = \frac{m_1^2 + m_2^2}{2}
- \frac{m_1^2 m_2^2}{m_1^2 - m_2^2}\ln\frac{m_1^2}{m_2^2}.
\label{eq:Floop}
\end{equation}
The $S$ and $U$ parameters receive only logarithmically suppressed 
corrections from the IDM that are negligible for the compressed spectra 
considered here. The PDG 2022 two-sigma bound $|\Delta T| < 
0.12$~\cite{ParticleDataGroup:2022pth} is the operative constraint; our 
benchmark spectrum with $\Delta m = 3~\mathrm{GeV}$ gives 
$\Delta T = -3.6 \times 10^{-4}$, four orders of magnitude below this 
bound (see Figure~\ref{fig:charged}).

\subsection{Benchmark parameters}

The model has nine free parameters beyond the SM:
$\{\mu_\phi^2, \lambda_\phi, \lambda_{H\phi}, \mu_\eta^2, \lambda_\eta,
\lambda_3, \lambda_4, \lambda_5, \lambda_{\phi\eta}\}$. In practice,
$\mu_\eta^2$ is traded for $m_{\eta^0_R}$ via Eq.~\eqref{eq:mueta}, and
$\mu_\phi^2$ is traded for $m_\phi$ via Eq.~\eqref{eq:mphi}. We fix
$\lambda_\eta = 0.10$ throughout, as the IDM self-coupling does not enter
any of the observables computed in this paper at leading order.

\paragraph{Singlet sector.}
The three free parameters of the singlet sector are $\{m_\phi,
\lambda_{H\phi}, \lambda_\phi\}$. Their values are constrained by two
independent physical requirements. First, the condition $\mu_\phi^2 < 0$,
which reads $\lambda_{H\phi} > 2m_\phi^2/v^2$ from
Eq.~\eqref{eq:lHphi_condition}, is necessary for the singlet to develop a
non-zero vacuum expectation value at high temperature and trigger Step~I.
Second, the Espinosa--Quiros condition for the two-step mechanism to produce
a strongly first-order electroweak transition at Step~II
requires~\cite{Espinosa:1996qw}
\begin{equation}
\lambda_{H\phi} > \sqrt{2\lambda_H\lambda_\phi},
\label{eq:EQ}
\end{equation}
which ensures that the barrier between the singlet-broken and
electroweak-broken phases is sufficiently large. Together, these two
conditions define a viable region in the $(\lambda_{H\phi}, m_\phi)$ plane
for each value of $\lambda_\phi$; this region is mapped in
Appendix~\ref{app:paramscan}. The three benchmark points are chosen to
sample this region at representative values of $m_\phi$ spanning the range
$70$--$110~\mathrm{GeV}$, with $\lambda_{H\phi}$ and $\lambda_\phi$ chosen
to satisfy both conditions while remaining well within the perturbative
regime $\lambda_{H\phi}, \lambda_\phi < 4\pi$.

\paragraph{IDM sector.}
The IDM internal couplings are fixed to
\begin{equation}
\lambda_3 = 0.08, \qquad \lambda_4 = 0.02, \qquad \lambda_5 = -0.05,
\label{eq:IDMcouplings}
\end{equation}
giving $\lambda_{H\eta}^{\rm eff} = 0.05$, $\Delta m \equiv
m_{\eta^\pm} - m_{\eta^0_R} = 3.0~\mathrm{GeV}$, and $\delta m \equiv
m_{\eta^0_I} - m_{\eta^0_R} = 9.8~\mathrm{GeV}$. The sign $\lambda_5 < 0$
is required to ensure $\eta^0_R$ is the lightest inert state and therefore
the stable dark matter candidate. The hierarchy $|\lambda_4|, |\lambda_5|
\ll \lambda_3$ keeps the charged-neutral and pseudoscalar-scalar mass
splittings small, producing a compressed spectrum with $\Delta m \ll m_W$.
The results of this paper are insensitive to the precise values of
$\lambda_4$ and $\lambda_5$ as long as this hierarchy holds and
$\lambda_5 < 0$; a discussion of the dependence on these parameters is
given in Appendix~\ref{app:paramscan}.

The benchmark IDM mass $m_{\eta^0_R} = 150~\mathrm{GeV}$ is chosen to
satisfy three requirements simultaneously. It lies above the Higgs
resonance at $m_h/2 = 62.5~\mathrm{GeV}$, so the IDM does not contribute
to the Higgs invisible width. It lies above the LEP2 lower bound on
charged scalar pair production, $m_{\eta^\pm} \gtrsim
70~\mathrm{GeV}$~\cite{Lundstrom:2008ai}. And at $\lambda_{H\eta}^{\rm
eff} \sim 0.05$--$0.20$, it falls in the parameter region where the IDM
can account for the deficit in the total relic density left by the
underabundant singlet component, as detailed in
Section~\ref{subsec:relic}.

The inter-sector coupling is fixed to 
$\lambda_{\phi\eta} = 0.05$. At this value the conversion channel 
$\phi\phi \leftrightarrow \eta^0_R\eta^0_R$ has no significant effect 
on either relic abundance: for the benchmark mass ordering 
$m_\phi < m_{\eta^0_R}$ the forward process is endothermic and 
Boltzmann-suppressed at singlet freeze-out, while the reverse process 
is suppressed relative to the gauge-mediated annihilation of the inert 
doublet, as quantified in Section~\ref{subsec:relic}. The two relic 
densities may therefore be computed independently. The value 
$\lambda_{\phi\eta} = 0.05$ is chosen small enough to avoid 
significant radiative corrections to the inert mass spectrum while 
remaining representative of a generic renormalisable coupling between 
the two dark sectors.

\begin{table}[h]
\centering
\renewcommand{\arraystretch}{1.3}
\begin{tabular}{lccc}
\hline\hline
Parameter & BP1 & BP2 & BP3 \\
\hline
$m_\phi~[\mathrm{GeV}]$ & 70 & 85 & 110 \\
$\lambda_{H\phi}$ & 0.50 & 0.60 & 0.80 \\
$\lambda_\phi$ & 0.40 & 0.50 & 0.80 \\
$\mu_\phi^2~[\mathrm{GeV}^2]$ & $-10229$ & $-10930$ & $-12106$ \\
\hline
$m_{\eta^0_R}~[\mathrm{GeV}]$ & \multicolumn{3}{c}{150 (fixed)} \\
$m_{\eta^0_I}~[\mathrm{GeV}]$ & \multicolumn{3}{c}{159.8} \\
$m_{\eta^\pm}~[\mathrm{GeV}]$ & \multicolumn{3}{c}{153.0} \\
$\Delta m~[\mathrm{GeV}]$ & \multicolumn{3}{c}{3.0} \\
$\lambda_{H\eta}^{\rm eff}$ & \multicolumn{3}{c}{0.05} \\
\hline\hline
\end{tabular}
\caption{Benchmark parameter sets. The IDM parameters are identical across
all BPs. The values of $\mu_\phi^2$ follow from inverting
Eq.~\eqref{eq:mphi}, $\mu_\phi^2 = m_\phi^2 - \lambda_{H\phi}v^2/2$, and
are negative for every benchmark, as required for Step~I of the phase
transition. The IDM soft mass is $\mu_\eta^2 = m_{\eta^0_R}^2 -
\lambda_{H\eta}^{\rm eff}\, v^2/2 = 20{,}987~\mathrm{GeV}^2$ for all BPs.}
\label{tab:benchmarks}
\end{table}

A comment on the fixed inert-doublet parameters is in 
order. The inert sector plays no dynamical role in the two-step 
transition, entering the effective potential only through loop 
contributions along the Higgs direction, whose quantitative effect is 
the few-GeV shift in $T_2$ shown in the inset of 
Figure~\ref{fig:vev}. The results of this paper are therefore 
insensitive to the internal IDM couplings, and we fix 
$\lambda_3, \lambda_4, \lambda_5$ to a representative compressed 
spectrum rather than scanning them; the residual dependence is 
quantified in Appendix~\ref{app:paramscan}, where the observables shift 
by less than $5\%$ across the ranges considered. The mass 
$m_{\eta^0_R} = 150~\mathrm{GeV}$ is likewise a representative choice, 
selected so that the inert states lie above the Higgs resonance and 
the LEP2 bound while retaining the compressed spectrum, 
$\Delta m \ll m_W$, whose collider and electroweak-precision 
phenomenology we discuss in Section~\ref{subsec:charged}. It should not 
be read as a prediction of the relic-density requirement: as noted in 
Section~\ref{subsec:relic}, a complete inert-doublet abundance 
calculation including gauge-mediated annihilation and coannihilation 
shifts the mass at which the deficit is saturated upward, into the 
familiar high-mass IDM regime. Since the no-go result of 
Section~\ref{subsec:nogo} concerns the singlet sector alone and is 
independent of the inert-doublet mass, this does not affect any 
conclusion of this work.

\section{Electroweak Phase Transition}
\label{sec:ewpt}

\subsection{Finite-temperature effective potential}
\label{subsec:veff}

The thermal history of the scalar sector is governed by the 
finite-temperature effective potential, which we compute at one loop 
using the Arnold--Espinosa daisy resummation 
scheme~\cite{Arnold:1992rz, Espinosa:1992gq}. Because the 
inert doublet never develops a vacuum expectation value, 
$\mathbb{Z}_2^{(2)}$ remains unbroken at all temperatures and the phase 
transition dynamics take place entirely in the two-dimensional field 
space $(h,\phi)$. The effective potential is
\begin{equation}
V_{\rm eff}(h,\phi,T) = V_0(h,\phi) + V_{\rm CW}(h,\phi) 
+ V_T(h,\phi,T) + V_{\rm daisy}(h,\phi,T),
\label{eq:Veff}
\end{equation}
where $V_0$ is the tree-level potential of Eq.~\eqref{eq:potential}, 
$V_{\rm CW}$ is the one-loop Coleman--Weinberg 
potential~\cite{Coleman:1973jx}, $V_T$ is the one-loop thermal 
correction, and $V_{\rm daisy}$ resums the ring (daisy) diagrams that 
become infrared-divergent at high temperature. Although the inert 
doublet does not define a field direction in Eq.~\eqref{eq:Veff}, it 
is fully included in the calculation: the loop sums in $V_{\rm CW}$, 
$V_T$, and $V_{\rm daisy}$ run over all scalar, vector, and fermion 
degrees of freedom of the model, including $\eta^0_R$, $\eta^0_I$, and 
$\eta^\pm$, whose field-dependent masses 
$M_{\eta_i}^2(h,\phi)$ depend on 
the Higgs background through the portal couplings of 
Eq.~\eqref{eq:Veta} and on the singlet background through 
$\lambda_{\phi\eta}$, as given in Appendix~\ref{app:veff}, and 
therefore contribute to the thermal corrections 
along the Higgs direction. The quantitative impact of the inert doublet 
on the Step~II nucleation temperature is shown in the inset of 
Figure~\ref{fig:vev}. The full expressions for all contributions are 
collected in Appendix~\ref{app:veff}; here we summarise the structure 
relevant to the two-step mechanism.

All finite-temperature quantities in this work are 
computed in the Landau gauge, in which the Goldstone and gauge-boson 
field-dependent masses take the standard forms of 
Appendix~\ref{app:veff}. The one-loop effective potential carries a 
residual gauge dependence through the Nielsen 
identity~\cite{Nielsen:1975fs, Fukuda:1975di}, and it is worth 
separating which of our results this systematic can affect. The 
boundaries of the viable region shown in Figure~\ref{fig:paramscan}, 
namely $\mu_\phi^2 = 0$ and the Espinosa--Quiros condition 
$\lambda_{H\phi} = \sqrt{2\lambda_H\lambda_\phi}$, are relations among 
Lagrangian parameters and are therefore gauge-independent; they do not 
move. Gauge dependence enters only through the thermal quantities 
evaluated at the scan points, $T_c^{(1,2)}$, $T_N^{(1,2)}$, and 
$\xi_H$. Comparisons of Landau gauge with general $R_\xi$ gauges and 
with gauge-invariant treatments find that $v_c/T_c$ in 
singlet-extended models typically shifts at the $10$--$30\%$ 
level~\cite{Patel:2011th, Wainwright:2011qy}, with critical 
temperatures moving by a few GeV. Such a shift would displace the 
$\xi_H = 1$ contour within the viable region while leaving its outer 
boundary intact. Since the surviving points span 
$\xi_H = 1.04$--$3.24$, only those lying within a few percent of the 
threshold could migrate across it under a change of gauge; the bulk of 
the region, including all three benchmarks at $\xi_H = 2.35$, $2.14$, 
and $1.69$, is unaffected. The gravitational wave predictions are 
correspondingly robust, since a few-GeV shift in $T_N$ propagates to 
the peak frequency at the percent level, far below the factor-of-two 
$\beta/H$ systematic already reflected in the uncertainty bands of 
Figure~\ref{fig:gw}. A fully gauge-invariant treatment along the lines 
of Ref.~\cite{Patel:2011th} is left to future work.

The Coleman--Weinberg potential in the $\overline{\rm MS}$ scheme is
\begin{equation}
V_{\rm CW}(h,\phi) = \sum_i \frac{n_i}{64\pi^2}\,M_i^4(h,\phi)
\left[\ln\frac{M_i^2(h,\phi)}{\mu^2} - c_i\right],
\label{eq:VCW}
\end{equation}
where the sum runs over all scalar, vector, and fermion degrees of 
freedom, $n_i$ are the corresponding multiplicities (positive for 
bosons, negative for fermions), $M_i^2(h,\phi)$ are the 
field-dependent masses, $\mu$ is the renormalisation scale taken to 
be $\mu = v$, and $c_i = 3/2$ for scalars and fermions, $c_i = 5/6$ 
for gauge bosons~\cite{Quiros:1999jp}. The field-dependent masses for 
all degrees of freedom, including the inert doublet, are listed in 
Appendix~\ref{app:veff}.

The one-loop thermal correction is
\begin{equation}
V_T(h,\phi,T) = \frac{T^4}{2\pi^2}\sum_i n_i\,
J_{B/F}\!\left(\frac{M_i^2(h,\phi)}{T^2}\right),
\label{eq:VT}
\end{equation}
where the thermal functions are
\begin{equation}
J_{B/F}(x^2) = \int_0^\infty dk\,k^2\,
\ln\left(1 \mp e^{-\sqrt{k^2+x^2}}\right),
\label{eq:JBF}
\end{equation}
with the upper (lower) sign for bosons (fermions). In the high-$T$ 
expansion these reduce to known analytic 
forms~\cite{Quiros:1999jp, Dolan:1973qd}, but we evaluate them 
numerically throughout to maintain accuracy across the full temperature 
range relevant to the two-step transition.

The daisy resummation accounts for the breakdown of perturbation theory 
for the bosonic zero Matsubara modes at high temperature. Following the 
Arnold--Espinosa prescription~\cite{Arnold:1992rz}, we replace the 
tree-level bosonic masses in $V_T$ by thermally corrected masses:
\begin{equation}
M_i^2(h,\phi) \to \bar{M}_i^2(h,\phi,T) = M_i^2(h,\phi) + \Pi_i(T),
\label{eq:daisy}
\end{equation}
and add the ring-diagram contribution
\begin{equation}
V_{\rm daisy}(h,\phi,T) = -\frac{T}{12\pi}\sum_{i\in{\rm bosons}}
n_i\left[\bar{M}_i^3(h,\phi,T) - M_i^3(h,\phi)\right].
\label{eq:Vdaisy}
\end{equation}
The thermal self-energies $\Pi_i(T)$ for the Higgs, singlet, inert 
doublet, and gauge boson sectors are listed in 
Appendix~\ref{app:veff}. The inert doublet fields contribute to 
the thermal masses of both the Higgs and the singlet through their 
portal couplings, and are included in the resummation.

\subsection{The two-step mechanism}
\label{subsec:twostep}
The phase structure at a given temperature is determined by the 
stationary points of the effective potential,
\begin{equation}
\frac{\partial V_{\rm eff}(h,\phi,T)}{\partial h} = 0, 
\qquad
\frac{\partial V_{\rm eff}(h,\phi,T)}{\partial \phi} = 0,
\label{eq:minima}
\end{equation}
subject to the positivity of the Hessian, 
$\partial^2 V_{\rm eff}/\partial\varphi_i\partial\varphi_j \succ 0$, 
which selects the minima among the solutions. Three minima are 
relevant to the two-step history: the symmetric phase 
$\mathcal{P}_{\rm sym} = (0,0)$, the intermediate phase 
$\mathcal{P}_{\rm int} = (0, \phi_c(T))$ of 
Eq.~\eqref{eq:intermediate}, and the electroweak phase 
$\mathcal{P}_{\rm EW} = (v_c(T), 0)$. A first-order transition between 
two of these occurs when they become degenerate at a critical 
temperature and are separated by a barrier. The two critical 
temperatures are defined by
\begin{align}
V_{\rm eff}(\mathcal{P}_{\rm sym}, T_c^{(1)}) 
&= V_{\rm eff}(\mathcal{P}_{\rm int}, T_c^{(1)}),
\label{eq:Tc1} \\
V_{\rm eff}(\mathcal{P}_{\rm int}, T_c^{(2)}) 
&= V_{\rm eff}(\mathcal{P}_{\rm EW}, T_c^{(2)}),
\label{eq:Tc2}
\end{align}
with $T_c^{(1)} > T_c^{(2)}$. A genuine two-step history requires this 
ordering together with the condition that the intermediate phase be 
the global minimum throughout the window 
$T_c^{(2)} < T < T_c^{(1)}$; when instead $\mathcal{P}_{\rm EW}$ 
becomes degenerate with $\mathcal{P}_{\rm sym}$ before 
$\mathcal{P}_{\rm int}$ is ever the global minimum, the singlet phase 
is bypassed and the transition proceeds directly to the electroweak 
vacuum in a single step. The latter case is discarded by the scan 
described in Section~\ref{subsec:scan}.

At high temperature $T \gg v$, both the Higgs and singlet fields are 
stabilised at the origin by thermal corrections and the system resides 
in the fully symmetric phase $\langle h\rangle = \langle\phi\rangle = 0$. 
As the universe cools, the effective potential develops structure that 
drives a sequence of two first-order transitions before the system 
reaches the zero-temperature electroweak vacuum at 
$\langle h\rangle = v$, $\langle\phi\rangle = 0$.

\paragraph{Step I: Singlet first-order transition.}
Below a temperature $T_1$, a second minimum develops along the singlet 
direction at $(\langle h\rangle, \langle\phi\rangle) = (0, \phi_c)$ 
with $\phi_c \neq 0$. This occurs because $\mu_\phi^2 < 0$, so the 
tree-level singlet potential already favours $\phi \neq 0$, while the 
thermal corrections generate a barrier separating the symmetric and 
singlet-broken phases that shrinks as $T$ decreases. At the nucleation 
temperature $T_N^{(1)}$, the tunnelling rate through this barrier 
exceeds the Hubble rate and the universe undergoes a first-order 
transition into the intermediate phase
\begin{equation}
\mathcal{P}_{\rm int}:\quad 
\langle h\rangle = 0,\quad \langle\phi\rangle = \phi_c(T) \neq 0.
\label{eq:intermediate}
\end{equation}
During this phase, $\mathbb{Z}_2^{(1)}$ is spontaneously broken and 
$\phi$ is technically unstable against decay to SM particles through 
the non-zero VEV. However, since $T_N^{(1)} \sim 200$--$270~\mathrm{GeV}$ 
is well above the $\phi$ freeze-out temperature 
$T_{\rm fo}^\phi \sim m_\phi/20 \sim 3.5$--$5.5~\mathrm{GeV}$, the 
singlet remains in full thermal equilibrium throughout the intermediate 
phase. The abundance of $\phi$ is therefore unaffected by the transient 
symmetry breaking, and $\mathbb{Z}_2^{(1)}$ is restored at Step~II, 
ensuring the stability of $\phi$ as a dark matter candidate at late 
times.

\paragraph{Step II: Strongly first-order electroweak transition.}
While the system resides in $\mathcal{P}_{\rm int}$, the non-zero 
singlet VEV $\phi_c(T)$ modifies the Higgs effective potential through 
the portal coupling $\lambda_{H\phi}\phi_c^2|H|^2/2$. This generates 
an additional positive contribution to the effective Higgs mass in the 
intermediate phase, which delays the electroweak transition and creates 
a barrier between $\mathcal{P}_{\rm int}$ and the electroweak-broken 
phase $\mathcal{P}_{\rm EW}: \langle h\rangle = v_c \neq 0$, 
$\langle\phi\rangle = 0$. At the nucleation temperature $T_N^{(2)}$, 
the system tunnels into $\mathcal{P}_{\rm EW}$ in a strongly 
first-order transition. The Espinosa--Quiros condition,
\begin{equation}
\lambda_{H\phi} > \sqrt{2\lambda_H\lambda_\phi},
\label{eq:EQ_main}
\end{equation}
is a necessary condition for this mechanism to generate a sufficiently 
strong barrier~\cite{Espinosa:1996qw, Profumo:2007wc}. As shown in 
Appendix~\ref{app:paramscan}, the benchmark points are chosen to 
satisfy this condition. A crucial structural result of the two-step 
mechanism is that $\xi_H \equiv v_c/T_c \geq 1$ is automatically 
satisfied for all parameter points that undergo the two-step sequence and it is not an additional constraint but a consequence of the intermediate-phase dynamics.

\begin{figure}[t]
\centering
\includegraphics[width=\textwidth]{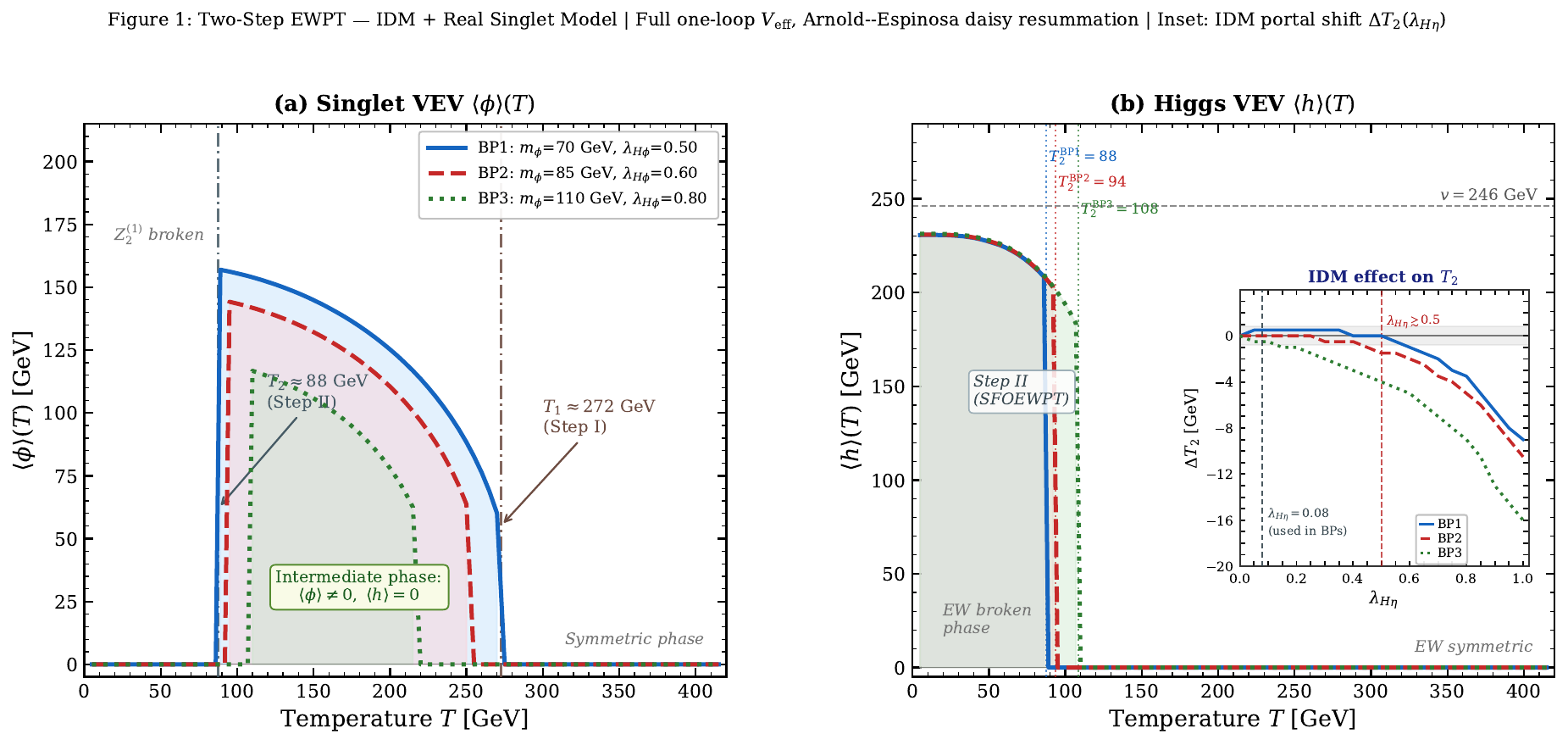}
\caption{VEV evolution through the two-step transition at the three 
benchmarks. Left: singlet VEV $\langle\phi\rangle(T)$, non-zero at 
Step~I and returning to zero at Step~II, establishing the intermediate 
phase $\mathcal{P}_{\rm int}$ (shaded). Right: Higgs VEV 
$\langle h\rangle(T)$, whose sharp rise at Step~II signals the strongly 
first-order electroweak transition. The labelled 
temperatures, $T_1 \simeq 272~\mathrm{GeV}$ and the inset values 
$88$, $94$, $108~\mathrm{GeV}$, are critical temperatures 
$T_c^{(1,2)}$; the nucleation temperatures of 
Table~\ref{tab:ptparams} lie below them by roughly $8$ and 
$2~\mathrm{GeV}$ respectively. The inset shows the shift 
$\Delta T_2(\lambda_{H\eta})$ from the IDM portal coupling at the 
benchmark $\lambda_{H\eta} = 0.08$ (dashed line); for 
$\lambda_{H\eta} \lesssim 0.5$ it stays below $5~\mathrm{GeV}$ and does 
not qualitatively affect the dynamics.}
\label{fig:vev}
\end{figure}

Our numerical strategy is as follows. The phase structure which includes the 
location and depth of all minima of Eq.~\eqref{eq:Veff} as a function 
of temperature, the critical temperatures $T_c^{(1,2)}$, and the VEV 
trajectories $\langle\phi\rangle(T)$ and $\langle h\rangle(T)$ is 
traced numerically with the \texttt{CosmoTransitions} 
package~\cite{Wainwright:2011kj}. The nucleation temperatures 
$T_N^{(1,2)}$ and the inverse durations $\beta/H$ are then estimated 
in the thin-wall approximation, as detailed in 
Section~\ref{subsec:ptparams}. The VEV evolution through both steps 
for the three benchmark points is shown in Figure~\ref{fig:vev} 
(left panel for the singlet, right panel for the Higgs). The tunnelling 
from $\mathcal{P}_{\rm int}$ to $\mathcal{P}_{\rm EW}$ at Step~II 
produces the strongly first-order electroweak transition, visible as 
the sharp discontinuity in $\langle h\rangle(T)$ at $T = T_N^{(2)}$.

\subsection{Phase transition parameters}
\label{subsec:ptparams}

The efficiency of gravitational wave production and the viability of 
electroweak baryogenesis are both controlled by three parameters 
evaluated at the nucleation temperature $T_N$ of each 
step~\cite{Espinosa:2010hh, Caprini:2019egz}:

\begin{itemize}
\item The phase transition strength 
$\alpha_N \equiv \Delta\rho/\rho_{\rm rad}|_{T=T_N}$, where 
$\Delta\rho$ is the latent heat released during the transition and 
$\rho_{\rm rad} = \pi^2 g_* T_N^4/30$ is the radiation energy density. 
For $\alpha_N \lesssim 0.1$ the transition is in the sound-wave 
dominated regime.

\item The inverse duration $\beta/H \equiv T_N\,(d S_3/dT)/S_3|_{T=T_N}$, 
where $S_3$ is the three-dimensional Euclidean bounce action. Larger 
$\beta/H$ corresponds to a shorter transition and a higher peak 
frequency for the gravitational wave spectrum.

\item The transition strength parameter $\xi_H \equiv v_c/T_c$, where 
$v_c$ is the Higgs VEV at the critical temperature $T_c$ of Step~II. 
The requirement $\xi_H \geq 1$ is the standard criterion for 
electroweak baryogenesis~\cite{Rubakov:1996vz, Morrissey:2012db}.
\end{itemize}

Since Step~I does not involve the electroweak symmetry, $\xi_H$ is 
defined only for Step~II. In the present model, $\xi_H \geq 1$ is 
automatically satisfied for all two-step parameter points, as noted 
in Section~\ref{subsec:twostep}. The values of $\beta/H$ quoted in this paper are obtained from a finite-difference 
derivative of the numerically computed bounce action, evaluating the temperature 
derivative of the action over a finite window, rather than from the full numerical bounce solution. A complete bounce computation across the 
scan is computationally demanding and is deferred to a future 
analysis; the associated shift in $\beta/H$ affects the peak frequency 
and amplitude of the gravitational wave spectrum at the 
$\mathcal{O}(1)$ level but does not alter the qualitative conclusion 
regarding DECIGO reach established in Section~\ref{sec:gw}.
The phase transition parameters for the three benchmark points are 
summarised in Table~\ref{tab:ptparams}.

\begin{table}[h]
\centering
\renewcommand{\arraystretch}{1.3}
\begin{tabular}{lccccccc}
\hline\hline
 & \multicolumn{3}{c}{Step I (Singlet FOPT)} & \multicolumn{4}{c}{Step II (SFOEWPT)} \\
\cmidrule(lr){2-4}\cmidrule(lr){5-8}
BP & $T_N^{(1)}$ & $\alpha_N^{(1)}$ & $\beta/H^{(1)}$ & $T_N^{(2)}$ & $\alpha_N^{(2)}$ & $\beta/H^{(2)}$ & $\xi_H$ \\
 & [GeV] & & & [GeV] & & & \\
\hline
BP1 & 264 & $1.5\times10^{-4}$ & 4572 & 86 & 0.039 & 6011 & 2.35 \\
BP2 & 245 & $3.0\times10^{-4}$ & 4266 & 92 & 0.033 & 6431 & 2.14 \\
BP3 & 210 & $8.0\times10^{-4}$ & 3504 & 106 & 0.021 & 7512 & 1.69 \\
\hline\hline
\end{tabular}
\caption{Phase transition parameters for the three benchmark points. 
$T_N^{(1,2)}$ are the nucleation temperatures for Step~I and Step~II 
respectively, $\alpha_N^{(1,2)}$ the transition strength parameters, and 
$\beta/H^{(1,2)}$ the inverse durations. Higgs vacuum expectation values $v_c = 207$, $201$, $182~\mathrm{GeV}$ for BP1, BP2, BP3 
respectively; $\xi_H \geq 1$ for all three.}
\label{tab:ptparams}
\end{table}

Several features of Table~\ref{tab:ptparams} deserve comment. 
Step~I is characterised by a very small $\alpha_N^{(1)} \sim 10^{-4}$--
$10^{-3}$, indicating a weak transition with little latent heat release 
is consistent with the singlet FOPT being a relatively mild event that 
does not significantly reheat the plasma. Step~II, by contrast, has 
$\alpha_N^{(2)} \sim 0.02$--$0.04$, placing it in the sound-wave 
dominated regime ($\alpha_N \ll 1$) where the dominant contribution 
to the gravitational wave signal comes from acoustic waves in the 
plasma. The strength parameter $\xi_H$ decreases from BP1 to BP3 
as $m_\phi$ increases, reflecting the weaker portal coupling needed 
to maintain the two-step sequence at larger singlet masses.

\subsection{Parameter scan and results}
\label{subsec:scan}

To identify the region of parameter space capable of realising a genuine
two-step electroweak phase transition, we perform a systematic scan over
the singlet-sector parameters,
\begin{equation}
m_\phi \in [42,154]~\mathrm{GeV}, \qquad
\lambda_{H\phi}\in[0.30,1.12], \qquad
\lambda_\phi\in[0.20,1.00].
\label{eq:scangrid}
\end{equation}
For each parameter point, the finite-temperature effective potential of
Eq.~\eqref{eq:Veff} is evaluated using
\texttt{CosmoTransitions}~\cite{Wainwright:2011kj} together with the
Arnold--Espinosa daisy resummation. Parameter points that do not satisfy
the necessary condition $\mu_\phi^2<0$ or violate the
Espinosa--Quiros criterion (Eq.~\eqref{eq:EQ_main}) are discarded before
the thermal evolution is analysed. The remaining points are then tested
for a complete two-step thermal history terminating in a strongly
first-order electroweak phase transition satisfying
$\xi_H \equiv v_c/T_c \geq 1$.

\begin{figure}[t]
\centering
\includegraphics[width=\textwidth]{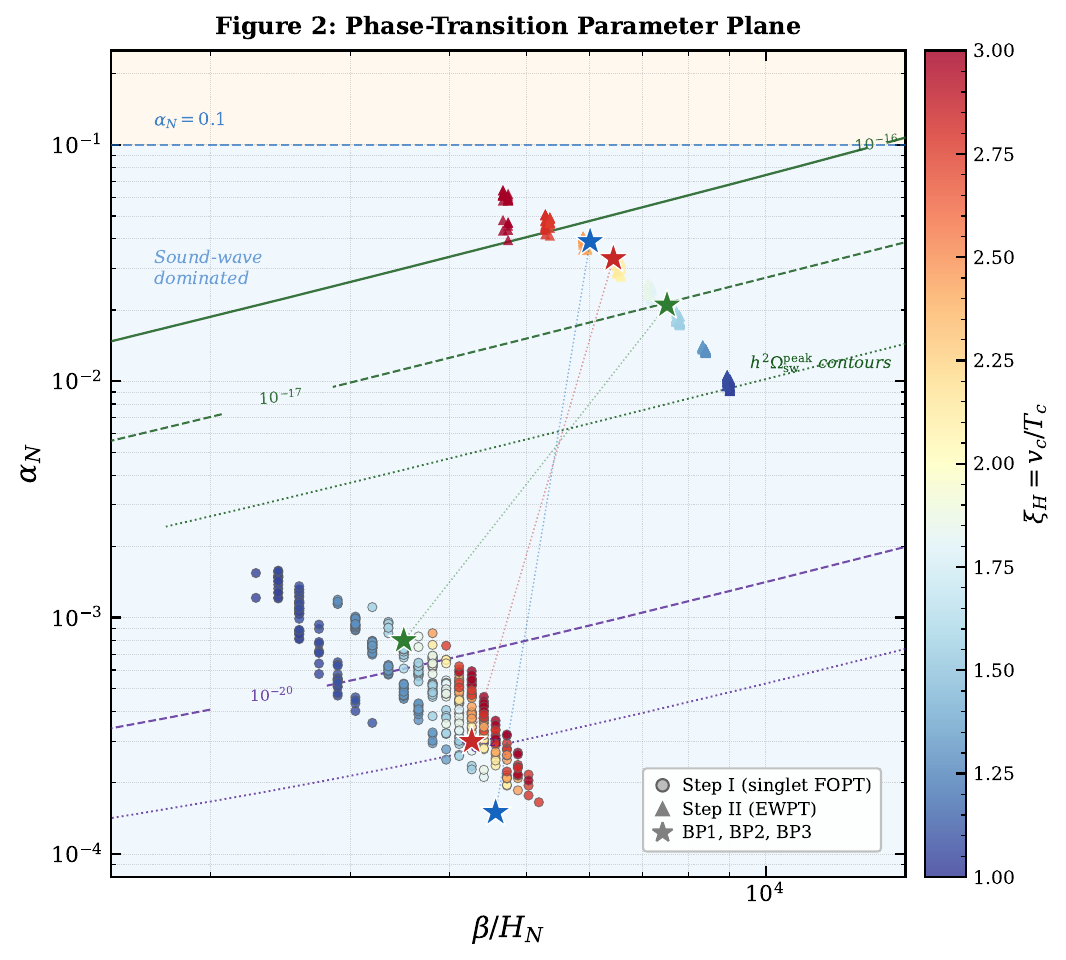}
\caption{Phase-transition parameter plane $(\alpha_N,\beta/H)$ for all
parameter points exhibiting a successful two-step electroweak phase
transition. Step~I (circles) corresponds to the singlet transition,
whereas Step~II (triangles) denotes the electroweak transition. The
points are coloured according to the strength of the electroweak phase
transition, $\xi_H=v_c/T_c$, while the benchmark points BP1, BP2 and
BP3 are highlighted by stars. Step~I is characterised by
$\alpha_N^{(1)}\sim10^{-4}$--$10^{-3}$ and
$\beta/H^{(1)}\sim2\times10^3$--$5\times10^3$, whereas Step~II exhibits
$\alpha_N^{(2)}\sim0.01$--$0.10$ with
$\beta/H^{(2)}\sim4\times10^3$--$9\times10^3$, placing the resulting
gravitational-wave signal in the sound-wave dominated regime.}
\label{fig:ptplane}
\end{figure}

Only a small subset of the scanned parameter space satisfies all of the
above requirements. The overwhelming majority of parameter points fail
because the singlet never develops an intermediate vacuum, the barrier
between the intermediate and electroweak phases is insufficient to
support a first-order transition, or the intermediate phase is bypassed
entirely in favour of a single-step crossover. The surviving solutions,
however, are not isolated fine-tuned points. Instead, they populate a
continuous and highly correlated region of parameter space, indicating
that once the necessary theoretical conditions are fulfilled, the
two-step mechanism is a robust feature of the model rather than an
accidental consequence of isolated benchmark choices. The benchmark
points used throughout this work are selected from this viable region
and are indicated in Fig.~\ref{fig:ptplane}.

Figure~\ref{fig:ptplane} reveals several generic features of the viable
parameter space. First, every surviving parameter point satisfies
$\xi_H>1$ without imposing this condition as an independent scan
criterion, demonstrating that a sufficiently strong electroweak phase
transition arises naturally once the required two-step thermal history
is realised. Second, the electroweak transition strength is typically
$\alpha_N^{(2)}\simeq0.01$--$0.10$, indicating that the gravitational-wave
signal is dominated by long-lived acoustic waves in the plasma rather
than by bubble-wall collisions. Finally, the corresponding inverse
duration falls in the range
$\beta/H^{(2)}\simeq(4$--$9)\times10^3$, implying peak frequencies in the
decihertz band where future space-based interferometers such as DECIGO
achieve their highest sensitivity. These features persist across the entire viable parameter space and are therefore representative of the model rather than peculiar to the benchmark points adopted in this work.

\section{Gravitational Wave Signal}
\label{sec:gw}

\subsection{Nucleation dynamics}
\label{subsec:nucleation}

A first-order transition proceeds through the stochastic nucleation of 
bubbles of the true vacuum inside the metastable phase, and it is the 
violent expansion, collision, and subsequent plasma dynamics of these 
bubbles that source a stochastic gravitational wave background. In the 
present model this happens twice: once when the singlet tunnels out of 
the fully symmetric phase at $T_N^{(1)}$, and again when the Higgs 
field escapes the intermediate phase $\mathcal{P}_{\rm int}$ at 
$T_N^{(2)}$. Both events take place at cosmic times of order 
$10^{-11}~\mathrm{s}$, far earlier than Big Bang Nucleosynthesis, so 
neither is constrained by light-element abundances. Their only relic 
is the gravitational radiation computed below.

The decay of the metastable vacuum is controlled by the 
finite-temperature tunnelling rate per unit 
volume~\cite{Linde:1981zj, Grojean:2006bp},
\begin{equation}\label{eq:tunnelling-prob}
\Gamma(T) \approx T^4 \left( \frac{S_E(T)}{2\pi T} \right)^{3/2} 
e^{-S_E(T)/T},
\end{equation}
in which the exponent is set by the $O(3)$-symmetric bounce 
action~\cite{Linde:1981zj}
\begin{equation}\label{eq:bounce-action}
S_E(T) = 4\pi \int_{0}^{\infty} r^2\, dr \left[ 
\frac{1}{2} \left( \frac{d\vec{\varphi}}{dr} \right)^2 
+ V_{\rm eff}(\vec{\varphi}, T) \right],
\qquad \vec{\varphi} = (h, \phi).
\end{equation}
The critical bubble profile extremising this action obeys
\begin{equation}\label{eq:phivariationr}
\frac{d^2 \vec{\varphi}}{dr^2} + \frac{2}{r} \frac{d\vec{\varphi}}{dr} 
= \nabla_{\vec{\varphi}}\, V_{\rm eff}(\vec{\varphi}, T),
\end{equation}
together with regularity at the bubble centre and asymptotic approach 
to the false vacuum~\cite{Affleck:1980ac, Linde:1977mm}:
\begin{equation}\label{bc:12}
\frac{d\vec{\varphi}}{dr}\bigg|_{r = 0} = 0,
\qquad
\lim_{r \to \infty} \vec{\varphi}(r) = \vec{\varphi}_{\rm false}.
\end{equation}
For Step~I the false vacuum is the origin of field space. For Step~II 
it is $\mathcal{P}_{\rm int}$, so the bounce trajectory at the second 
step genuinely explores the two-dimensional $(h,\phi)$ plane rather 
than a single field direction. We obtain the bounce solutions with 
\texttt{CosmoTransitions}~\cite{Wainwright:2011kj} and identify the 
nucleation temperature of each step from the standard criterion of 
one bubble per horizon volume, $S_E(T_N)/T_N \simeq 140$.

A comment is in order regarding topological defects. 
Because Step~I spontaneously breaks the discrete $\mathbb{Z}_2^{(1)}$, 
domain walls form between causally disconnected regions that select 
opposite signs of $\langle\phi\rangle = \pm\phi_c$. The scalar 
potential of Eq.~\eqref{eq:potential} is exactly even in $\phi$, so 
the two vacua are exactly degenerate and no volume pressure biases one 
against the other; the walls are therefore not removed by a bias. They 
are instead removed by the second step of the transition itself. The 
intermediate phase is transient: at $T_N^{(2)}$ the singlet returns to 
the origin, $\langle\phi\rangle \to 0$, restoring $\mathbb{Z}_2^{(1)}$ 
and erasing the distinction between the $\pm\phi_c$ domains, so the 
wall network cannot survive past Step~II. That Step~II completes across 
the horizon, and hence erases the walls everywhere rather than only 
locally, follows from the nucleation criterion 
$S_E(T_N^{(2)})/T_N^{(2)} \simeq 140$ used throughout: in a 
radiation-dominated background this corresponds to at least one bubble 
nucleated per Hubble volume, which is sufficient for percolation and 
completion of the transition. Its cosmological impact 
during the short interval $T_N^{(2)} < T < T_N^{(1)}$ is negligible: 
with wall tension $\sigma \sim \phi_c^3 \sim 10^{7}~\mathrm{GeV}^3$ at 
$T \sim 100~\mathrm{GeV}$, the scaling-regime energy fraction 
$\rho_{\rm wall}/\rho_{\rm rad} \sim \sigma H/\rho_{\rm rad} 
\sim 10^{-17}$ is minute, and wall domination would occur only at 
$t \sim M_{\rm Pl}^2/\sigma \sim 10^{5}~\mathrm{s}$, some sixteen 
orders of magnitude after the intermediate phase has ended at 
$t \sim 10^{-11}~\mathrm{s}$. The Zel'dovich--Kobzarev--Okun bound on 
stable domain walls~\cite{Zeldovich:1974uw} does not apply, since it 
governs walls of a symmetry that remains broken, whereas here 
$\mathbb{Z}_2^{(1)}$ is restored. The one caveat is that a transient wall network can seed 
inhomogeneous nucleation of Step~II~\cite{Blasi:2022woz, 
Agrawal:2023cgp}, effectively lowering the effective $\beta/H^{(2)}$. 
For seeded transitions the enhancement is typically at the 
$\mathcal{O}(10$--$50\%)$ level in $\beta/H$ in the most affected 
region, which would raise the Step~II amplitude by a comparable factor 
through the $(\beta/H)^{-2}$ scaling of the sound-wave peak; we do not 
include this enhancement, so our Step~II amplitudes are conservative.

The strength and the pace of each transition are encoded in the two 
quantities introduced in Section~\ref{subsec:ptparams}. The energy 
budget parameter $\alpha_N$ compares the trace-anomaly difference 
between the phases to the radiation energy 
density~\cite{Kamionkowski:1993fg, Kehayias:2009tn},
\begin{equation}\label{eq:delrho}
\Delta \rho = \left[ V_{\rm eff} 
- T\, \partial_T V_{\rm eff} \right]_{\vec{\varphi}_{\rm false},\,T_N}
- \left[ V_{\rm eff} 
- T\, \partial_T V_{\rm eff} \right]_{\vec{\varphi}_{\rm true},\,T_N},
\end{equation}
while the inverse duration in Hubble units follows from the 
temperature dependence of the action near nucleation,
\begin{equation}\label{eq:betaP}
\frac{\beta}{H_N} = T_N\, \frac{d}{dT} 
\left( \frac{S_E(T)}{T} \right) \bigg|_{T_N}.
\end{equation}
The benchmark values of both quantities, for both steps, were listed 
in Table~\ref{tab:ptparams}. Physically, the hierarchy between the two 
steps is already visible there. Step~I liberates almost no latent heat, 
with $\alpha_N^{(1)} \lesssim 10^{-3}$, whereas Step~II releases enough 
to drive detectable acoustic waves. Both steps complete rapidly, with 
$\beta/H_N \gtrsim 3500$, a direct consequence of the steep 
temperature dependence of the bounce action in the two-step geometry.

\subsection{The spectrum and its three components}
\label{subsec:gwspectrum}

Three physically distinct processes convert the liberated vacuum 
energy into gravitational radiation: the collision of the bubble walls 
themselves, the long-lived acoustic waves the walls launch into the 
plasma, and the magnetohydrodynamic turbulence that develops once the 
acoustic phase breaks down. To a good approximation the three 
contributions add incoherently~\cite{Caprini:2015zlo},
\begin{equation}\label{eq:GWTotal}
\Omega_{\rm GW}h^2 \approx \Omega_{\rm col} h^2 + \Omega_{\rm sw} h^2 
+ \Omega_{\rm tur} h^2,
\end{equation}
with \( h = H_0/(100~\mathrm{km\,s^{-1}\,Mpc^{-1}}) 
\)~\cite{DES:2017txv}. For transitions with $\alpha_N \ll 1$ and 
relativistic walls, the acoustic contribution dominates the peak, the 
turbulence term controls the high-frequency tail, and wall collisions 
are numerically negligible. We nevertheless retain all three.

The collision term, evaluated in the envelope 
approximation~\cite{Jinno:2016vai}, reads
\begin{equation}\label{eq:GWcoldetails}
\Omega_{\rm col} h^2 = 1.67 \times 10^{-5} 
\left(\frac{\beta}{H_N} \right)^{-2} 
\left( \frac{\kappa_c\, \alpha_N}{1 + \alpha_N} \right)^2 
\left( \frac{100}{g_{*}} \right)^{1/3} 
\left( \frac{0.11\, v^3_w}{0.42 + v^2_w} \right) 
\frac{3.8 \left( f/f_{\rm col} \right)^{2.8}}
{1 + 2.8 \left( f/f_{\rm col} \right)^{3.8}},
\end{equation}
where $g_* = 106.75$~\cite{Husdal:2016haj} and the fraction of vacuum 
energy deposited in the walls 
is~\cite{Kamionkowski:1993fg, Borah:2023zsb}
\begin{equation}\label{eq:kcfac}
\kappa_c = \frac{0.715\, \alpha_N 
+ \frac{4}{27} \sqrt{\frac{3 \alpha_N}{2}}}{1 + 0.715\, \alpha_N}.
\end{equation}
Its present-day peak frequency is
\begin{equation}\label{eq:PF1}
f_{\rm col} = 16.5 \times 10^{-6} 
\left( \frac{f_{*}}{\beta} \right) 
\left( \frac{\beta}{H_N} \right) 
\left( \frac{T_N}{100~\mathrm{GeV}} \right) 
\left( \frac{g_{*}}{100} \right)^{1/6} \mathrm{Hz},
\qquad
\frac{f_{*}}{\beta} = \frac{0.62}{1.8 - 0.1 v_w + v^2_w},
\end{equation}
with the fit of Ref.~\cite{Jinno:2016vai}. For the wall velocity we adopt the Chapman--Jouguet detonation value,
\begin{equation}\label{eq:vJ}
v_w = v_J = \frac{1/\sqrt{3} + \sqrt{\alpha_N^2 + 2\alpha_N/3}}
{1 + \alpha_N},
\end{equation}
which yields $v_J \simeq 0.72$, $0.71$, and $0.68$ for BP1, BP2, and 
BP3 respectively. This choice is not a convenience but a consistency 
requirement: the acoustic efficiency factor of 
Eq.~\eqref{eq:kappasw} is the numerical fit of 
Ref.~\cite{Espinosa:2010hh} for Jouguet detonations, so pairing it 
with any other wall velocity would combine incompatible hydrodynamic 
solutions.\footnote{A genuinely luminal wall, $v_w \to 1$, is not 
realised for the transition strengths encountered here. Runaway 
requires the vacuum pressure to exceed the plasma friction at all 
Lorentz factors, and the next-to-leading-order friction from 
transition radiation of soft gauge bosons grows with the wall 
$\gamma$-factor~\cite{Bodeker:2017cim}, guaranteeing a terminal 
velocity for $\alpha_N \ll 1$. In the $v_w \to 1$ limit the 
appropriate efficiency would instead be 
$\kappa \simeq \alpha_N/(0.73 + 0.083\sqrt{\alpha_N} + \alpha_N)$, 
a factor of $\simeq 3.9$ smaller than 
Eq.~\eqref{eq:kappasw} at $\alpha_N = 0.03$; the corresponding spectra 
would be suppressed by a factor of $\simeq 15$ in amplitude.}

We note one tension that the detonation choice makes explicit. 
Electroweak baryogenesis conventionally favours subsonic walls, since 
efficient charge transport requires CP-violating currents to diffuse 
into the plasma ahead of the wall, whereas a Jouguet detonation is 
supersonic by construction.\footnote{Two remarks are in 
order. First, the wall velocity is not the binding obstacle to 
baryogenesis in this model: as noted in Section~\ref{sec:intro}, the 
scalar potential of Eq.~\eqref{eq:potential} contains no CP-violating 
phase beyond that of the Standard Model, so no asymmetry is generated 
regardless of $v_w$. The discussion here therefore concerns what would 
be required were a CP source added, for instance through a complex 
extension of the singlet sector or higher-dimensional operators 
coupling the scalars to fermions. Second, the supersonic-wall tension 
is generic to scenarios linking a strong gravitational wave signal to 
baryogenesis and admits several partial resolutions in the literature. 
Hybrid solutions, in which a deflagration front is followed by a 
rarefaction wave, retain a region of non-zero fluid velocity ahead of 
the wall in which diffusion can operate~\cite{No:2011fi, 
Cline:2020jre}. Local friction from the next-to-leading-order 
transition radiation that guarantees a terminal velocity in the first 
place also acts to keep realistic walls closer to the sound speed than 
the Jouguet value in much of parameter space~\cite{Bodeker:2017cim, 
Hoche:2020ysm}. Transport analyses that solve the Boltzmann equations 
without assuming a subsonic wall find that the baryon asymmetry, while 
suppressed relative to the deflagration case, need not vanish for 
supersonic walls~\cite{Dorsch:2021ubz, Cline:2021iff}. A 
self-consistent determination of $v_w$ for the present model, solving 
the friction and transport equations together, is beyond the scope of 
this work and would be a prerequisite for any quantitative 
baryogenesis claim.} The sensitivity of our results to 
this choice is mild. The peak frequency scales as $1/v_w$, while the 
acoustic amplitude scales as $v_w^2$, since the explicit factor of 
$v_w$ in Eq.~\eqref{eq:GWswpart} is accompanied by a second power 
entering through $R_N \propto v_w$ in the lifetime factor 
$\Upsilon$. Replacing $v_J \simeq 0.7$ by a subsonic 
$v_w \simeq 0.6$ therefore lowers the amplitude by roughly $27\%$ and 
raises the peak frequency by roughly $17\%$, shifts that are small 
compared with the systematic uncertainties displayed in 
Figure~\ref{fig:gw} and that leave the detectability of the strongest 
benchmarks unchanged.

Sound waves are the dominant source. Numerical simulations of the 
coupled field-fluid system~\cite{Hindmarsh:2013xza, 
Hindmarsh:2016lnk, Hindmarsh:2017gnf} yield
\begin{equation}\label{eq:GWswpart}
\Omega_{\rm sw} h^2 = 2.65 \times 10^{-6}\; \Upsilon(\tau_{\rm sw}) 
\left( \frac{\beta}{H_N} \right)^{-1} v_w 
\left( \frac{\kappa_{\rm sw}\, \alpha_N}{1 + \alpha_N} \right)^2 
\left( \frac{100}{g_{*}} \right)^{1/3} 
\left( \frac{f}{f_{\rm sw}} \right)^3 
\left[ \frac{7}{4 + 3 \left( f/f_{\rm sw} \right)^2} \right]^{7/2},
\end{equation}
where the acoustic efficiency 
is~\cite{Kamionkowski:1993fg, Borah:2025hpo}
\begin{equation}\label{eq:kappasw}
\kappa_{\rm sw} = \frac{\sqrt{\alpha_N}}{0.135 
+ \sqrt{0.98 + \alpha_N}},
\end{equation}
and the peak sits at
\begin{equation}\label{eq:PF2}
f_{\rm sw} = 1.9 \times 10^{-5} \left( \frac{1}{v_w} \right) 
\left( \frac{\beta}{H_N} \right) 
\left( \frac{T_N}{100~\mathrm{GeV}} \right) 
\left( \frac{g_{*}}{100} \right)^{1/6} \mathrm{Hz}.
\end{equation}
The prefactor $\Upsilon$ corrects for the fact that the acoustic 
source does not radiate for a full Hubble 
time~\cite{Hindmarsh:2017gnf}:
\begin{equation}\label{eq:swtimepart}
\Upsilon(\tau_{\rm sw}) = 1 - \frac{1}{\sqrt{1 + 2\, \tau_{\rm sw} 
H_{*}}},
\qquad
\tau_{\rm sw} \approx \frac{R_N}{\overline{U}_f},
\qquad
R_N = (8\pi)^{1/3}\, \frac{v_w}{\beta},
\qquad
\overline{U}_f = \sqrt{\tfrac{3}{4}\kappa_{\rm sw}\alpha_N}.
\end{equation}
This suppression is far from cosmetic in our scenario. Rapid 
transitions produce small bubbles. With $\beta/H_N^{(2)} \sim 
6000$--$7500$ the mean separation $R_N$ is minuscule, the shocks 
develop early, and $\tau_{\rm sw}H_* \sim 10^{-3}$--$10^{-2}$, so 
$\Upsilon$ reduces the acoustic amplitude by two to three orders of 
magnitude relative to the naive long-lived-source estimate. 
Because $\Upsilon \propto (\beta/H_N)^{-1}$ in this 
short-source regime, the acoustic peak amplitude scales as 
$(\beta/H_N)^{-2}$ rather than $(\beta/H_N)^{-1}$, which is the 
scaling reflected in the systematic bands of 
Figure~\ref{fig:gw}. Any 
statement about detectability of the two-step scenario that omits 
this factor is unreliable, and all spectra shown in this paper 
include it.

Once the acoustic period ends, a fraction of the bulk kinetic energy 
cascades into MHD turbulence in the fully ionised 
plasma~\cite{Caprini:2009yp}:
\begin{equation}\label{eq:GWturpart}
\Omega_{\rm tur} h^2 = 3.35 \times 10^{-4} 
\left( \frac{\beta}{H_N} \right)^{-1} v_w 
\left( \frac{\kappa_{\rm tur}\, \alpha_N}{1 + \alpha_N} \right)^{3/2} 
\left( \frac{100}{g_{*}}\right)^{1/3} 
\frac{\left( f/f_{\rm tur} \right)^3}
{\left[ 1 + \left( f/f_{\rm tur} \right) \right]^{11/3} 
\left( 1 + 8\pi f/h_{*} \right)},
\end{equation}
with
\begin{equation}\label{eq:PF3}
f_{\rm tur} = 2.7 \times 10^{-5} \frac{1}{v_w} 
\left( \frac{\beta}{H_N} \right) 
\left( \frac{T_N}{100~\mathrm{GeV}} \right) 
\left( \frac{g_{*}}{100} \right)^{1/6} \mathrm{Hz},
\end{equation}
and 
\( h_{*} = 16.5\times 10^{-6}\, (T_N/100~\mathrm{GeV})\, 
(g_{*}/100)^{1/6}~\mathrm{Hz} \) 
the redshifted Hubble rate at production. Following common practice we 
take $\kappa_{\rm tur} = \epsilon\,\kappa_{\rm sw}$ with 
$\epsilon = 0.1$. The turbulence normalisation carries an $O(1)$ 
theoretical uncertainty that does not affect our conclusions, since 
the peak region is acoustic-dominated.

\subsection{Step~I versus Step~II and the decihertz window}
\label{subsec:gwresults}

\begin{figure}[t]
\centering
\includegraphics[width=\textwidth]{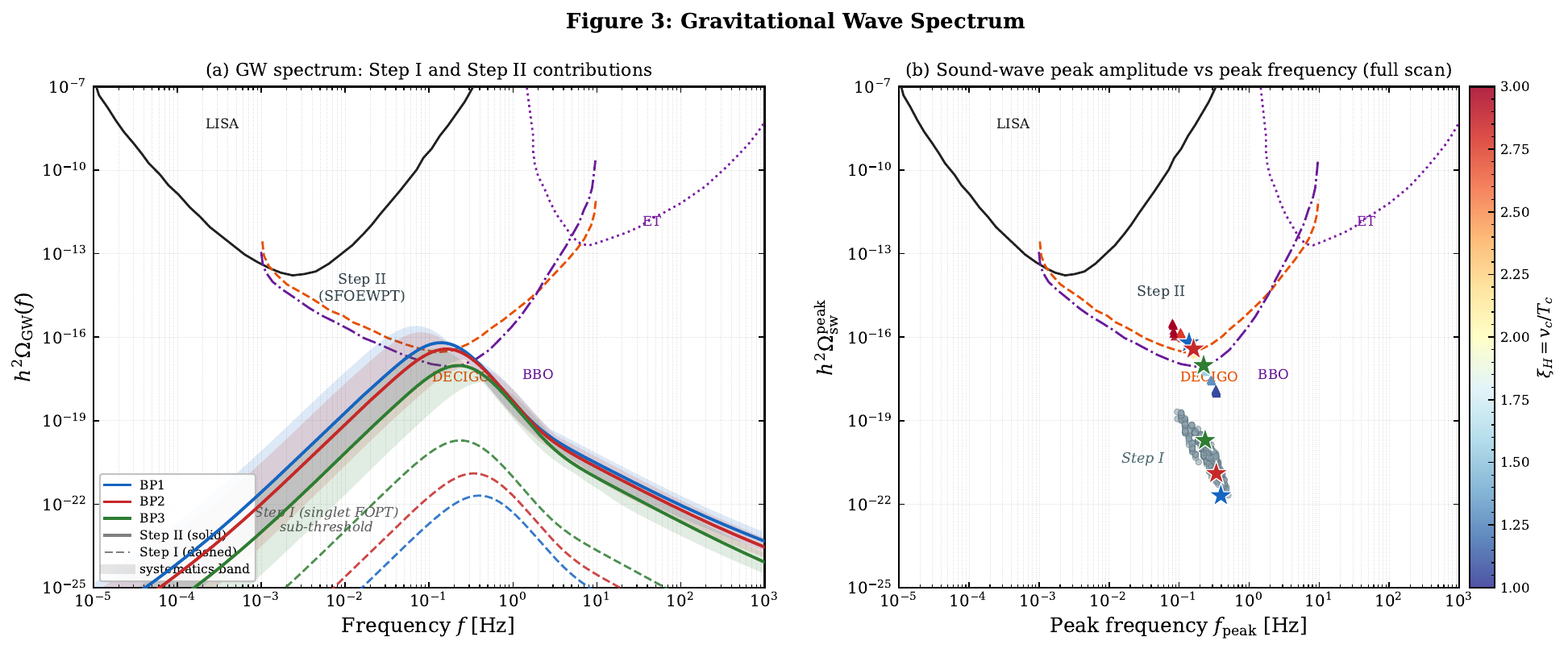}
\caption{Gravitational wave spectra 
$h^2\Omega_{\rm GW}(f)$ from the two-step transition. Left: full 
spectra at the three benchmark points, with Step~II (solid, SFOEWPT) 
and Step~I (dashed, singlet FOPT), against the power-law integrated 
sensitivities of LISA, DECIGO, BBO, and ET. Shaded bands show the 
envelope of the Step~II spectra under variation of $\beta/H$ by a 
factor of two in either direction and 
$\epsilon_{\rm tur}\in[0.05,0.15]$, the two dominant theoretical 
systematics discussed in the text. Right: sound-wave peak amplitude 
versus peak frequency for every point of the parameter scan, coloured 
by $\xi_H$, with Step~I shown for contrast. The Step~II peaks lie in 
the decihertz band, $f_{\rm peak}\simeq 0.08$--$0.34~\mathrm{Hz}$. 
The strongest transitions, including BP1 and BP2, lie above the DECIGO 
sensitivity; BP3 falls a factor of a few below DECIGO but within the 
reach of BBO. Step~I is sub-threshold for every planned instrument. 
The acoustic contribution dominates each peak, turbulence shapes the 
high-frequency tail, and wall collisions are negligible throughout.}
\label{fig:gw}
\end{figure}

Figure~\ref{fig:gw} displays the resulting spectra for both steps at 
the three benchmarks, overlaid on the power-law integrated sensitivity 
curves of LISA~\cite{LISA:2017pwj, Caprini:2015zlo}, 
DECIGO~\cite{Seto:2001qf, Kawamura:2011zz}, 
BBO~\cite{Harry:2006fi}, and ET~\cite{Punturo:2010zz}.

The two steps could hardly be more different observationally, despite 
originating from the same scalar potential. The singlet transition 
carries almost no latent heat, and its signal sits several orders of 
magnitude beneath every sensitivity curve. Step~I is the dynamical 
engine of the whole scenario, since without it there is no 
intermediate phase and no strong electroweak transition, yet it is 
invisible to gravitational wave astronomy. The information it carries 
reaches us only indirectly, through the properties it imprints on 
Step~II.

Step~II compensates. Its combination of $\alpha_N^{(2)} \sim 
0.02$--$0.04$ and $\beta/H_N^{(2)} \sim 6000$--$7500$ at 
the three benchmarks pushes the peak frequency to
\begin{equation}
f_{\rm peak}^{(2)} \simeq 0.14\text{--}0.22~\mathrm{Hz}
\quad(\text{benchmarks}),
\end{equation}
while across the full scan the Step~II peak spans 
$f_{\rm peak}^{(2)} \simeq 0.08$--$0.34~\mathrm{Hz}$, a band that is 
awkward for both LISA, whose sensitivity is optimised near the 
millihertz, and for ground-based interferometers, which operate above 
a few hertz. The decihertz gap is precisely the design territory of 
DECIGO and BBO. In this sense the large $\beta/H_N$ characteristic of 
the two-step mechanism, which suppresses the amplitude through both 
the explicit $1/(\beta/H_N)$ scaling and the lifetime factor 
$\Upsilon$, simultaneously relocates the signal into a frequency 
window where a dedicated instrument will operate with exquisite 
sensitivity. Whether a given parameter point is detectable is then a 
competition between these two effects.

Across the viable parameter points, the Step~II peak 
amplitudes computed with the suppression factor $\Upsilon$ included 
span about two and a half decades, from $h^2\Omega_{\rm sw}^{\rm peak} 
\sim 10^{-18}$ to $\sim 3\times 10^{-16}$. Roughly a quarter of the 
viable region, comprising the strongest transitions, produces signals 
lying above the DECIGO power-law integrated sensitivity curve; this 
set includes BP1 and BP2. The remaining points, including BP3, fall 
below the DECIGO curve by a factor of a few, but the deeper 
sensitivity of BBO reaches part of this population, BP3 among it. When 
the theoretical systematics indicated by the shaded bands in 
Figure~\ref{fig:gw} are folded in, the upper envelope of BP3 also 
crosses the DECIGO curve, so its detectability is marginal rather than 
excluded. This places the decihertz signal of Step~II at the centre 
of the phenomenology of the model. As Section~\ref{sec:dm} 
demonstrates, the LZ-2024 measurement excludes the singlet as a 
thermal relic precisely because the portal coupling required for the 
two-step mechanism yields a nucleon scattering cross section orders 
of magnitude above the experimental bound. Direct detection therefore 
ceases to function as a discovery channel for this scenario and 
instead acts as its sharpest constraint, while the gravitational wave 
signature, powered by the very same portal coupling, survives as the 
leading avenue for a positive observation. The collider signatures of 
the compressed inert-doublet spectrum, discussed in 
Section~\ref{subsec:charged}, offer a complementary but longer-term 
probe. We return to this interplay between exclusion and discovery in 
the conclusions.

\section{Dark Matter Phenomenology}
\label{sec:dm}

\subsection{Two-component relic density}
\label{subsec:relic}

Both dark sectors were in thermal contact with the SM plasma in the 
early universe: the singlet through the Higgs portal $\lambda_{H\phi}$ 
and the inert doublet through its gauge interactions and the portal 
couplings of Eq.~\eqref{eq:Veta}. Each species therefore undergoes 
thermal freeze-out, and the observed abundance must be shared,
\begin{equation}
\Omega_{\rm DM} h^2 = \Omega_\phi h^2 + \Omega_{\eta^0_R} h^2 
= 0.120,
\label{eq:relicsum}
\end{equation}
with the Planck value~\cite{Planck:2018vyg}. It is convenient to 
define the relic fractions
\begin{equation}
f_\phi = \frac{\Omega_\phi}{\Omega_{\rm DM}}, 
\qquad 
f_\eta = \frac{\Omega_{\eta^0_R}}{\Omega_{\rm DM}} = 1 - f_\phi,
\label{eq:fractions}
\end{equation}
which control the rescaling of direct detection signals in 
Section~\ref{subsec:dd}.

For each species the freeze-out abundance follows from the standard 
semi-analytic solution of the Boltzmann 
equation~\cite{Kolb:1990vq, Gondolo:1990dk},
\begin{equation}
\Omega_X h^2 \simeq 
\frac{1.07\times 10^9~\mathrm{GeV}^{-1}\, x_f}
{\sqrt{g_*}\, M_{\rm Pl}\, \langle\sigma v\rangle_X},
\qquad x_f = \frac{m_X}{T_{\rm fo}} \simeq 20,
\label{eq:relic}
\end{equation}
with $M_{\rm Pl} = 2.435\times 10^{18}~\mathrm{GeV}$ the reduced 
Planck mass and $g_* = 106.75$. For the singlet, the annihilation 
proceeds through $s$-channel Higgs exchange into $b\bar{b}$, 
$W^+W^-$, $ZZ$, and $t\bar{t}$, with 
$\langle\sigma v\rangle_\phi \propto \lambda_{H\phi}^2$.
%This treatment carries an $\mathcal{O}(50\%)$ uncertainty away from the Higgs resonance, well below the margin by which the exclusion of Section~\ref{subsec:nogo} holds there.
 In the vicinity of $m_\phi \simeq m_h/2$
the semi-analytic solution is less reliable, and the abundance at those points
is obtained from the thermally averaged cross section of Ref.~\cite{Gondolo:1990dk}.

The central feature of the two-component structure follows 
immediately. The two-step mechanism forces 
$\lambda_{H\phi} \gtrsim 0.32$ (Appendix~\ref{app:paramscan}), an 
annihilation cross section far above the thermal target, and the 
singlet freezes out underabundant across the entire viable region. 
At the benchmarks the analytic treatment gives
\begin{equation}
\Omega_\phi h^2 = 0.0047,\; 0.028,\; 0.065
\quad\Longrightarrow\quad
f_\phi = 0.039,\; 0.234,\; 0.539
\label{eq:BPfractions}
\end{equation}
for BP1, BP2, and BP3 respectively. The deficit must be carried by 
the inert doublet, requiring 
$\Omega_{\eta^0_R} h^2 = 0.115$, $0.092$, and $0.055$.

\begin{figure}[t]
\centering
\includegraphics[width=\textwidth]{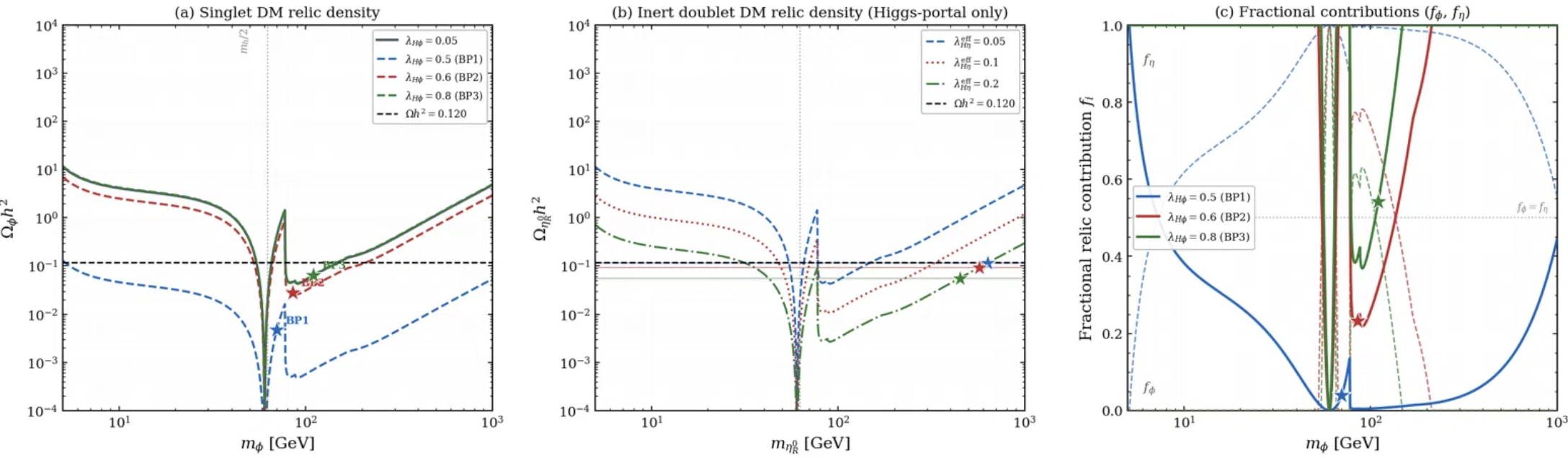}
\caption{Two-component relic density structure. Left: singlet 
abundance $\Omega_\phi h^2$ versus $m_\phi$ for representative 
$\lambda_{H\phi}$, benchmarks starred and the Higgs resonance dip at 
$m_h/2$. Middle: Higgs-portal contribution only to the 
inert-doublet abundance versus $m_{\eta^0_R}$ for three 
$\lambda_{H\eta}^{\rm eff}$; horizontal lines mark the required 
deficits $\Omega_{\eta^0_R} h^2 = 0.115, 0.092, 0.055$ at BP1--BP3, 
with stars where the $\lambda_{H\eta}^{\rm eff}=0.20$ curve meets 
each deficit. Right: the resulting relic fractions $f_\phi$ 
(solid) and $f_\eta$ (dashed), which sum to unity, with the benchmark 
values starred.}
\label{fig:relic}
\end{figure}

Figure~\ref{fig:relic} presents the relic structure. The left panel 
shows $\Omega_\phi h^2$ as a function of $m_\phi$ for representative 
portal couplings, exhibiting the Higgs resonance dip at 
$m_\phi = m_h/2$ and the systematic underabundance at the couplings 
demanded by the phase transition. The middle panel shows the 
Higgs-portal contribution to $\Omega_{\eta^0_R} h^2$ for 
$\lambda_{H\eta}^{\rm eff} = 0.05$, $0.10$, and $0.20$, together with 
the three target abundances; within this approximation the deficit is 
saturated at $m_{\eta^0_R} \simeq 186$--$425$, $190$--$387$, and 
$204$--$304~\mathrm{GeV}$ for the three benchmarks at 
$\lambda_{H\eta}^{\rm eff} = 0.20$. These mass ranges 
follow from the Higgs-portal channels alone and should not be read as 
predictions; as discussed below, the inclusion of gauge-mediated 
annihilation shifts them upward. The right panel shows the 
resulting fractional composition.

Two remarks accompany this analysis, and both act in the direction of 
strengthening rather than weakening the conclusions. First, the 
$\eta$ panel of Figure~\ref{fig:relic} retains only the Higgs-portal 
annihilation channels. The full inert doublet in the mass range of 
interest annihilates dominantly through electroweak gauge 
interactions, $\eta\eta \to W^+W^-$ and $ZZ$, together with 
coannihilation among the compressed states $\eta^0_R$, $\eta^0_I$, 
and $\eta^\pm$~\cite{LopezHonorez:2006gr, Belyaev:2016lok}. These 
channels are independent of $\lambda_{H\eta}^{\rm eff}$ and shift the 
mass at which the deficit is saturated toward the familiar high-mass 
IDM regime, while simultaneously permitting a small portal coupling. 
This decoupling of the relic density from 
$\lambda_{H\eta}^{\rm eff}$ is precisely what allows the $\eta$ 
component to evade direct detection, as discussed in 
Section~\ref{subsec:dd}.

Second, we address the inter-sector conversion channel 
$\phi\phi \leftrightarrow \eta^0_R\eta^0_R$ mediated by 
$\lambda_{\phi\eta}$. Because the inert doublet is heavier than the 
singlet throughout the benchmark set, 
$m_{\eta^0_R} = 150~\mathrm{GeV}$ against 
$m_\phi = 70$--$110~\mathrm{GeV}$, the forward process is endothermic 
and is closed at singlet freeze-out: producing a pair of inert states 
requires $2(m_{\eta^0_R} - m_\phi) = 160$, $130$, and 
$80~\mathrm{GeV}$ of thermal kinetic energy at BP1, BP2, and BP3, 
whereas $T_{\rm fo}^\phi \simeq m_\phi/20 \sim 3.5$--$5.5~
\mathrm{GeV}$. The corresponding Boltzmann factors, 
$e^{-2(m_{\eta^0_R} - m_\phi)/T_{\rm fo}^\phi} \sim 10^{-20}$, 
$10^{-14}$, and $10^{-7}$, render the channel entirely negligible, 
and the singlet abundance is fixed by Higgs-portal annihilation 
alone. The reverse process $\eta^0_R\eta^0_R \to \phi\phi$ is 
exothermic and remains open at inert-doublet freeze-out, but there it 
competes with gauge-mediated annihilation, from which it is 
suppressed by $\lambda_{\phi\eta}^2/g^4 \simeq 1.4\times 10^{-2}$; 
its effect on $\Omega_{\eta^0_R}$ is at the percent level. The two 
relic densities may therefore be computed independently, as we have 
done, and a coupled Boltzmann treatment would not measurably alter 
either $f_\phi$ or $f_\eta$ anywhere in the benchmark region. This 
conclusion relies on the mass ordering $m_\phi < m_{\eta^0_R}$; were 
the singlet the heavier state, the forward channel would open and a 
coupled treatment would be required.

Finally, we note that the transient breaking of $\mathbb{Z}_2^{(1)}$ 
during the intermediate phase has no effect on the relic abundance: 
as established in Section~\ref{subsec:twostep}, the singlet remains 
in full thermal equilibrium throughout the window 
$T_N^{(2)} < T < T_N^{(1)}$, its freeze-out occurring only at 
$T_{\rm fo}^\phi \sim m_\phi/20$, long after $\mathbb{Z}_2^{(1)}$ has 
been restored.

\subsection{Direct detection}
\label{subsec:dd}

The singlet scatters off nuclei through $t$-channel Higgs exchange 
with spin-independent cross 
section~\cite{Cline:2013gha}
\begin{equation}
\sigma_{\rm SI}^\phi = 
\frac{\lambda_{H\phi}^2\, f_N^2\, \mu_{\phi N}^2\, m_N^2}
{\pi\, m_h^4\, m_\phi^2},
\qquad 
\mu_{\phi N} = \frac{m_\phi m_N}{m_\phi + m_N},
\label{eq:sigmaSI}
\end{equation}
where $m_N = 0.939~\mathrm{GeV}$ and $f_N \simeq 0.30$ is the 
effective Higgs-nucleon coupling~\cite{Cline:2013gha, 
Hoferichter:2017olk}. Because the singlet constitutes only a fraction 
of the local dark matter density, the quantity constrained by 
experiment is the rescaled cross section
\begin{equation}
\tilde{\sigma}_{\rm SI}^\phi = f_\phi\, \sigma_{\rm SI}^\phi,
\label{eq:rescaled}
\end{equation}
under the standard assumption that the local composition tracks the 
cosmological one. This rescaling is maximally generous to the model: 
any local enhancement of the singlet fraction would only strengthen 
the exclusion below.

\begin{figure}[t]
\centering
\includegraphics[width=0.85\textwidth]{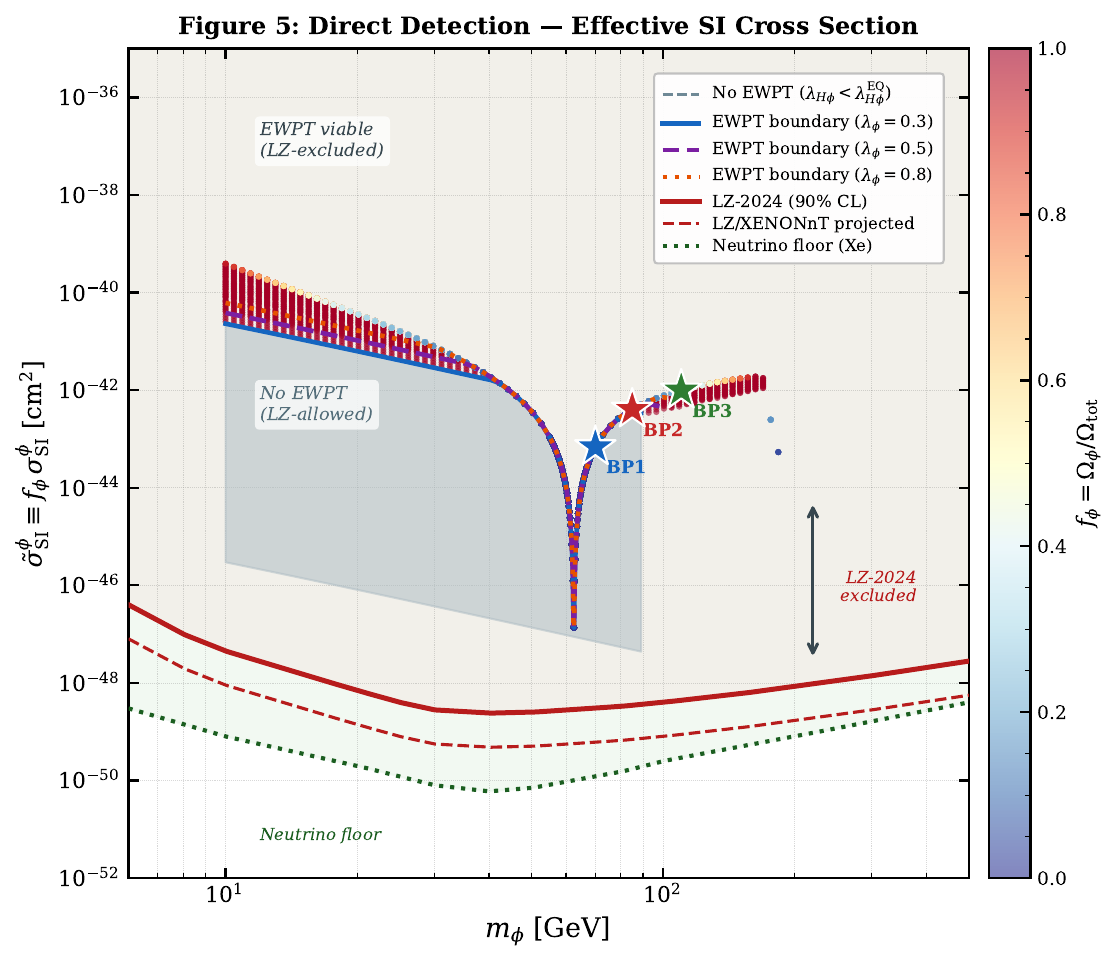}
\caption{Rescaled spin-independent cross section 
$\tilde{\sigma}_{\rm SI}^\phi = f_\phi\,\sigma_{\rm SI}^\phi$ versus 
$m_\phi$ for parameter points satisfying the analytic 
two-step conditions $\mu_\phi^2 < 0$ and 
$\lambda_{H\phi} > \sqrt{2\lambda_H\lambda_\phi}$, coloured by the 
relic fraction $f_\phi$. The solid red curve is the LZ-2024 exclusion 
limit~\cite{LZ:2024zvo}, the dashed curve the projected future 
sensitivity, and the shaded band at the bottom the neutrino fog. The 
grey band shows, for contrast, the coupling range allowed by LZ-2024, 
which supports no electroweak phase transition. Benchmark points are 
marked by stars. Every point lies well above the LZ-2024 
limit, by four to five orders of magnitude in cross section across 
the mass range.}
\label{fig:dd}
\end{figure}

Figure~\ref{fig:dd} confronts $\tilde{\sigma}_{\rm SI}^\phi$ with the 
LZ-2024 exclusion limit~\cite{LZ:2024zvo} across the full scan. 
The outcome is unambiguous. The phase transition demands
$\lambda_{H\phi} \gtrsim 0.32$, while LZ-2024 tolerates
$\lambda_{H\phi} \lesssim$ few$\times 10^{-4}$ over the mass range of the scan,
a gap of three orders of magnitude in the coupling and six in the unrescaled
cross section. The relic-fraction suppression recovers only a factor
$1/f_\phi$ of this, about an order and a half at BP1 and less at the heavier
benchmarks. The rescaled cross sections at the benchmarks exceed the LZ-2024
limit by factors of by factors of $2.3\times10^{4}$ (BP1), $1.2\times10^{5}$ (BP2), and
$2.1\times10^{5}$ (BP3). The margin
narrows near $m_\phi \simeq m_h/2$, where resonant annihilation drives
$f_\phi$ down, but the limit is exceeded there as well. Every parameter point
that completes the two-step transition is excluded.

The inert doublet component evades this fate for the structural 
reason identified in Section~\ref{subsec:relic}: its abundance is 
fixed by gauge interactions, so $\lambda_{H\eta}^{\rm eff}$, which 
alone controls $\sigma_{\rm SI}^{\eta}$ through the analogue of 
Eq.~\eqref{eq:sigmaSI}, is a free parameter from the point of view of 
the relic density and can sit comfortably below the LZ-2024 limit 
across the relevant mass range~\cite{Belyaev:2016lok}. Direct 
detection therefore bites only the component that drives the phase 
transition.

\subsection{Constraint map below the Higgs resonance}
\label{subsec:constraintmap}

The one region not covered by the scan of Figure~\ref{fig:dd} is the 
light window $m_\phi < m_h/2$, where the resonant enhancement of the 
annihilation cross section might be hoped to reconcile a small portal 
coupling with the observed abundance, and where the Higgs acquires 
the invisible decay channel $h \to \phi\phi$ with 
width
\begin{equation}
\Gamma_{h\to\phi\phi} = 
\frac{\lambda_{H\phi}^2 v^2}{32\pi\, m_h}
\sqrt{1 - \frac{4 m_\phi^2}{m_h^2}},
\label{eq:Gammainv}
\end{equation}
constrained by the ATLAS combination 
$\mathrm{BR}(h\to{\rm inv}) < 0.107$~\cite{ATLAS:2023tkt}.

\begin{figure}[t]
\centering
\includegraphics[width=0.85\textwidth]{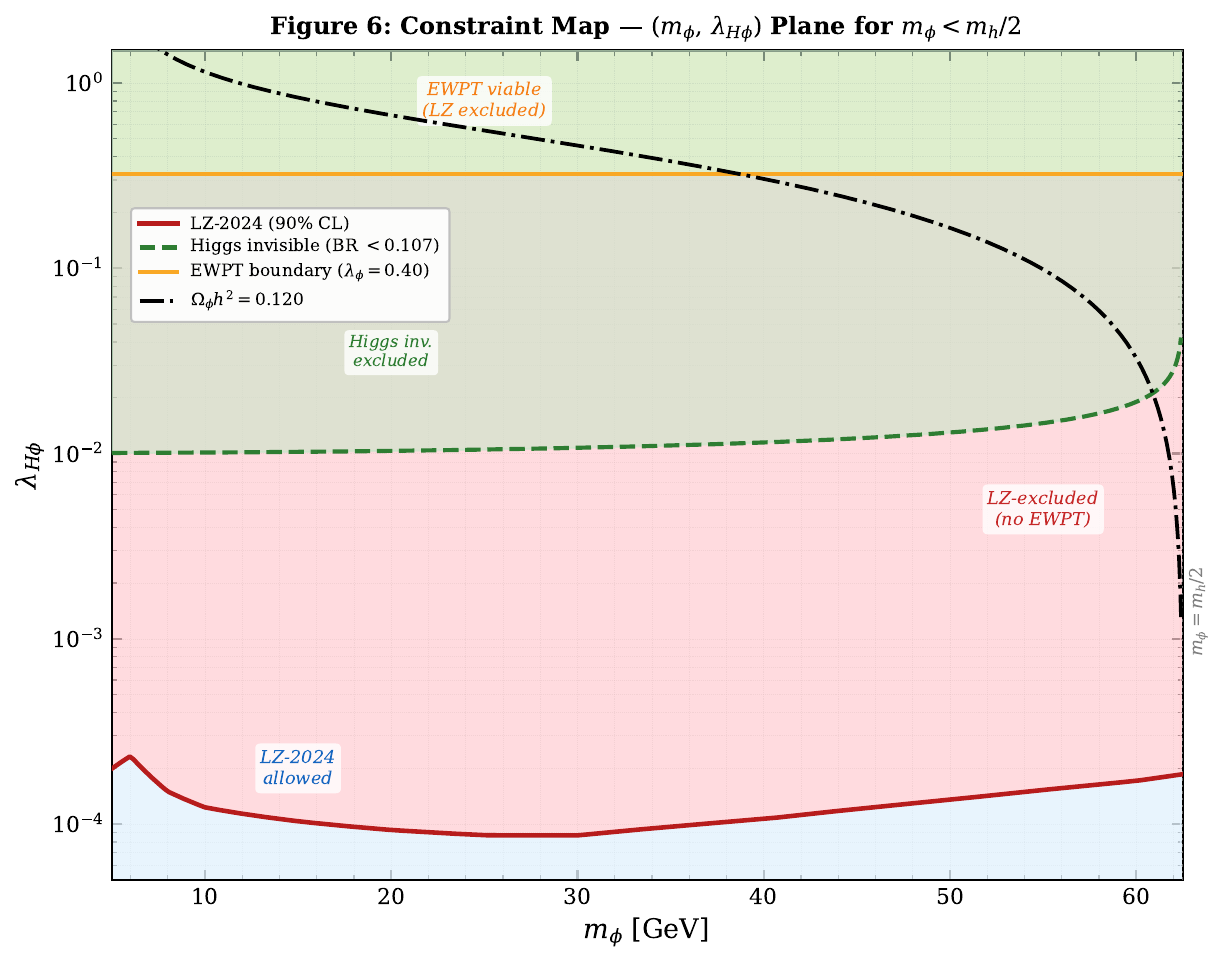}
\caption{Constraint map in the $(m_\phi, \lambda_{H\phi})$ plane for 
$m_\phi < m_h/2$. The red region is excluded by LZ-2024 after relic 
fraction rescaling, the green region by the Higgs invisible branching 
ratio $\mathrm{BR}(h\to{\rm inv}) < 0.107$~\cite{ATLAS:2023tkt}, and 
the yellow band marks the coupling range required by the two-step 
electroweak phase transition. The black curve is the contour along 
which the singlet alone reproduces the Planck abundance. No region of 
the plane satisfies the phase transition, relic density, and direct 
detection requirements simultaneously.}
\label{fig:constraints}
\end{figure}

Figure~\ref{fig:constraints} assembles all constraints in the 
$(m_\phi, \lambda_{H\phi})$ plane for this window. The LZ-2024 limit 
confines the coupling to a strip $\lambda_{H\phi} \lesssim 
2\times 10^{-4}$, the invisible-width bound excludes the region above 
$\lambda_{H\phi} \sim 10^{-2}$, and the two-step phase transition 
requires $\lambda_{H\phi} > 0.32$. The relic density contour, along 
which the singlet alone saturates Eq.~\eqref{eq:relicsum}, threads 
the LZ-excluded region except in the immediate vicinity of the 
resonance. The conclusion is visible at a glance: the EWPT-viable 
band and the LZ-allowed strip share no overlap anywhere in the 
window, and the resonance region rescues the relic density but not 
the phase transition, which fails there for every 
$\lambda_\phi$ in the scan range. The light window is closed.

\subsection{Charged scalar phenomenology}
\label{subsec:charged}

\begin{figure}[t]
\centering
\includegraphics[width=0.85\textwidth]{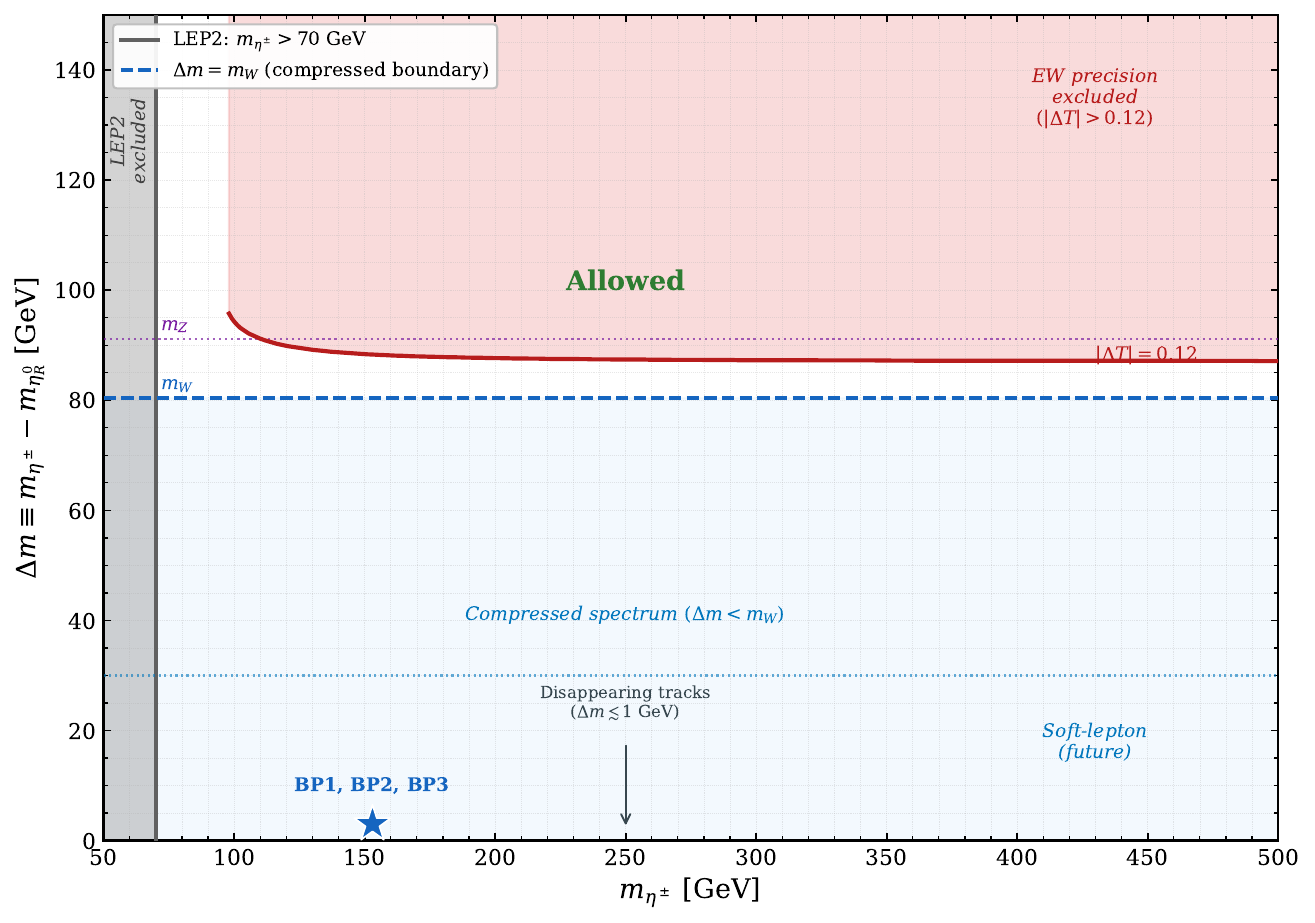}
\caption{Constraints in the $(m_{\eta^\pm}, \Delta m)$ plane. The 
grey region is excluded by LEP2 chargino-type searches, 
$m_{\eta^\pm} > 70~\mathrm{GeV}$~\cite{Lundstrom:2008ai}, and the red 
region by the oblique $T$ parameter, $|\Delta T| > 
0.12$~\cite{ParticleDataGroup:2022pth}. The dashed 
line $\Delta m = m_W$ separates on-shell from off-shell 
$\eta^\pm \to W^\pm \eta^0_R$ decays; below the dotted line the decay 
products are soft leptons, and for $\Delta m \lesssim 1~\mathrm{GeV}$ 
disappearing-track signatures apply. Both regimes are indicated as 
targets of future sensitivity rather than current exclusions, since 
no dedicated LHC recast for the inert doublet exists in this mass 
range. The star marks the benchmark spectrum, common to all three 
BPs.}
\label{fig:charged}
\end{figure}

The inert doublet contributes the only coloured-blind charged state 
of the model, and its phenomenology is fixed by the compressed 
spectrum of Table~\ref{tab:benchmarks}: 
$m_{\eta^\pm} = 153~\mathrm{GeV}$ with 
$\Delta m = m_{\eta^\pm} - m_{\eta^0_R} = 3~\mathrm{GeV}$. 
Figure~\ref{fig:charged} locates this point in the 
$(m_{\eta^\pm}, \Delta m)$ plane against the constraints that can be 
imposed without recourse to a dedicated collider recast.

Three statements are on firm footing. The benchmark clears the LEP2 
bound. Its oblique correction, evaluated from 
Eq.~\eqref{eq:DeltaT}, is $\Delta T = -3.6\times 10^{-4}$, four 
orders of magnitude inside the allowed range, a direct consequence of 
the near-degeneracy $\Delta m \ll m_{\eta^\pm}$. And with 
$\Delta m = 3~\mathrm{GeV}$, the decay 
$\eta^\pm \to \eta^0_R\, \ell^\pm \nu$ proceeds through a far 
off-shell $W$ and yields leptons with 
$p_T \sim \Delta m/2 \sim 1.5~\mathrm{GeV}$, below current trigger 
thresholds. The benchmark is thus invisible to existing LHC searches 
not by accident but by kinematics; conversely, it defines a concrete 
target for the soft-lepton and, at smaller splittings, 
disappearing-track programmes of the High-Luminosity 
LHC~\cite{Calibbi:2015nha}. We emphasise that the position of the 
benchmark in this plane is not tuned: as derived in 
Eqs.~\eqref{eq:Deltam} and~\eqref{eq:deltam}, the splitting follows 
from the hierarchy $|\lambda_4|, |\lambda_5| \ll \lambda_3$ adopted 
on independent grounds in Section~\ref{sec:model}, and the 
conclusions of this subsection are stable across the ranges of 
$\lambda_4$ and $\lambda_5$ quantified in 
Appendix~\ref{app:paramscan}. The charged scalar $\eta^\pm$ also contributes a loop 
correction to the diphoton rate $h \to \gamma\gamma$. For 
$m_{\eta^\pm} = 153~\mathrm{GeV}$ and 
$g_{h\eta^+\eta^-} = \lambda_3 v$ with $\lambda_3 = 0.08$, the shift 
in the signal strength is $|\Delta\mu_{\gamma\gamma}| \lesssim 1\%$, 
an order of magnitude inside the current ATLAS and CMS 
uncertainties~\cite{ATLAS:2022tnm, CMS:2021kom}, so the benchmark is 
unconstrained by Higgs signal-strength data.

\subsection{The combined no-go}
\label{subsec:nogo}

Assembling the results of this section yields a statement stronger 
than any individual constraint. The two-step transition requires 
$\lambda_{H\phi} > 0.32$; the observed relic density at such 
couplings leaves the singlet underabundant, which is consistent; but the rescaled direct detection cross section then exceeds the LZ-2024 bound
at every surviving parameter point, by four to five orders of magnitude across
most of the scan and by a narrower margin at the Higgs resonance, and the
sub-resonance window is independently closed by the 
combination of Figure~\ref{fig:constraints}. Within the thermal 
freeze-out paradigm and the standard local-density assumption, the 
real singlet of this model cannot simultaneously enable the strongly 
first-order electroweak phase transition and survive as a dark matter 
component: direct detection has converted the singlet from a 
candidate into pure phase-transition machinery. The dark matter of 
the model is the inert doublet alone, with $f_\eta \to 1$ enforced 
observationally rather than by construction, and the observable 
signatures of the singlet sector migrate entirely to the 
gravitational wave channel of Section~\ref{sec:gw} and, secondarily, 
to the compressed-spectrum collider signatures of 
Section~\ref{subsec:charged}. The conclusions quantify this division 
of labour.

\section{Conclusions}
\label{sec:conclusions}

We have analysed a minimal two-component dark sector, a real singlet 
$\phi$ and an inert doublet $\eta$ protected by independent 
$\mathbb{Z}_2$ symmetries, in which the two fields play sharply 
separated roles: the singlet restructures the thermal history of 
electroweak symmetry breaking, while the inert doublet supplies the 
dark matter. The analysis connects four observables, the phase 
transition, its gravitational wave relic, the relic abundance, and 
direct detection, within a single parameter space, and the resulting 
picture is unusually clean.

On the phase transition side, the condition $\mu_\phi^2 < 0$ together 
with the Espinosa--Quiros criterion 
$\lambda_{H\phi} > \sqrt{2\lambda_H\lambda_\phi}$ carves out a 
connected region of the $(m_\phi, \lambda_{H\phi}, \lambda_\phi)$ 
space in which the universe passes through a genuine two-step 
sequence: a singlet transition at $T_N^{(1)} \simeq 210$--$264~
\mathrm{GeV}$ into an intermediate phase with 
$\langle\phi\rangle \neq 0$, followed by a strongly first-order 
electroweak transition at $T_N^{(2)} \simeq 86$--$106~\mathrm{GeV}$. Every parameter point completing this sequence satisfies 
$\xi_H = v_c/T_c \geq 1$ automatically, with 
$\xi_H = 2.35, 2.14 ~\rm{and}~ 1.69$ at the three benchmarks; 
this washout-avoidance condition is a structural output 
of the mechanism rather than an imposed cut. We emphasise that 
$\xi_H \geq 1$ is necessary but not sufficient for electroweak 
baryogenesis: the scalar potential studied here contains no 
CP-violating phase beyond that of the Standard Model, so a complete baryogenesis calculation, with a CP source at the wall and a self-consistent wall velocity, lies outside the present scope.

The gravitational wave signal inherits the imprint of this structure. 
Step~I, with $\alpha_N^{(1)} \lesssim 10^{-3}$, is unobservably weak. 
Step~II, with $\alpha_N^{(2)} \simeq 0.02$--$0.04$ and 
$\beta/H^{(2)} \simeq (6$--$7.5)\times 10^3$ at the benchmarks, 
radiates through long-lived acoustic waves with Chapman--Jouguet wall 
velocities $v_J \simeq 0.68$--$0.72$, and its peak lands in the 
decihertz band near $0.14$--$0.22~\mathrm{Hz}$ for the 
benchmarks and $0.08$--$0.34~\mathrm{Hz}$ across the full scan, a 
window inaccessible to both LISA and ground-based interferometers but 
central to the DECIGO and BBO design. After the finite-lifetime 
suppression $\Upsilon \sim 10^{-3}$--$10^{-2}$ is included, an effect 
that reduces naive amplitude estimates by two to three orders of 
magnitude and cannot be omitted at these values of $\beta/H$, 
the strongest transitions in the viable region, roughly 
a quarter of the scan and including benchmarks BP1 and BP2, lie above 
the DECIGO power-law integrated sensitivity, while the remainder, BP3 
among them, fall a factor of a few below DECIGO but within the reach 
of BBO. Folding in the theoretical systematics on $\beta/H$ and the 
turbulence normalisation lifts the upper envelope of BP3 to the DECIGO 
curve, so its detectability is marginal rather than excluded.

On the dark matter side, the same portal coupling that powers the 
transition destroys the singlet as a thermal relic candidate. The 
floor $\lambda_{H\phi} \gtrsim 0.32$ leaves the singlet underabundant, 
with $f_\phi = 0.039$, $0.234$, and $0.539$ at the benchmarks in the 
thermal picture, and simultaneously drives the rescaled 
spin-independent cross section 
$\tilde{\sigma}_{\rm SI}^\phi = f_\phi\,\sigma_{\rm SI}^\phi$ above 
the LZ-2024 limit by factors of $2.2\times 10^5$, $1.2\times 10^6$, 
and $2.3\times 10^6$ respectively. The exclusion holds at every 
parameter point of the viable region,\footnote{The most recent LZ 
analysis~\cite{LZ:2026axp}, based on a $2.84$~tonne-year exposure with 
the nuclear recoil window extended to approximately $270~{\rm keV}$, 
targets effective field theory and inelastic interactions, whose recoil 
spectra extend to high energies. It reports one event at 
$248\pm23({\rm stat})\pm23({\rm sys})~{\rm keV}$, corresponding to a 
global significance of $2.6\sigma$ against the background-only 
hypothesis. The Higgs portal coupling of the singlet produces the 
standard spin-independent scalar interaction, whose recoil spectrum is 
concentrated at low energies, so this search does not strengthen the 
bound relevant here and its excess does not bear on the singlet. 
LZ-2024~\cite{LZ:2024zvo} therefore remains the operative bound 
throughout the viable region. A full-exposure limit at higher masses 
would only widen the margin of the exclusion.} and the light window 
$m_\phi < m_h/2$ is independently closed: there the Higgs invisible 
width bounds $\lambda_{H\phi} \lesssim 10^{-2}$, LZ-2024 confines it 
below a few$\,\times 10^{-4}$, and the transition requires 
$\lambda_{H\phi} > 0.32$, three demands with no common solution. 
Within standard freeze-out and the standard local density assumption, 
the singlet of this model is pure phase-transition machinery; the 
observed dark matter is the inert doublet, whose abundance is set by 
gauge interactions and whose small portal coupling keeps it below 
current direct detection sensitivity. Its compressed spectrum, 
$m_{\eta^\pm} = 153~\mathrm{GeV}$ with $\Delta m = 3~\mathrm{GeV}$ 
and $\Delta T = -3.6\times 10^{-4}$, evades LEP2 and electroweak 
precision constraints by construction and defines a target for 
soft-lepton searches at the High-Luminosity LHC.

The division of labour among experiments is therefore inverted 
relative to the usual Higgs-portal narrative. Direct detection, 
normally the discovery channel, here functions purely as the 
exclusion arm: it has already eliminated the singlet as dark matter 
while leaving the phase transition mechanism itself untouched, since 
the transition never required the singlet to be stable relic matter 
in the first place. Discovery capability migrates entirely to 
gravitational waves, where DECIGO and BBO would probe 
the surviving mechanism directly, with the compressed inert-doublet 
spectrum as a longer-term collider cross-check. A decihertz detection 
accompanied by continued direct detection null results and a 
soft-lepton excess would constitute a coherent, over-constrained 
signature of this scenario.

\appendix

\section{Finite-temperature effective potential: full expressions}
\label{app:veff}

This appendix collects every ingredient entering 
Eq.~\eqref{eq:Veff}: the tree-level background potential, the 
field-dependent mass spectrum including background mixing, the 
one-loop zero- and finite-temperature pieces, and the thermal 
self-energies used in the Arnold--Espinosa resummation.

\subsection{Tree-level background potential}
\label{app:tree}

In the background $\langle H\rangle = (0, h/\sqrt{2})^{\rm T}$, 
$\langle\phi\rangle = \phi$, $\langle\eta\rangle = 0$, the potential 
of Eq.~\eqref{eq:potential} reduces to
\begin{equation}
V_0(h,\phi) = -\frac{\mu_H^2}{2}h^2 + \frac{\lambda_H}{4}h^4 
+ \frac{\mu_\phi^2}{2}\phi^2 + \frac{\lambda_\phi}{4}\phi^4 
+ \frac{\lambda_{H\phi}}{4}h^2\phi^2 .
\label{eq:V0app}
\end{equation}
The inert doublet enters only through the loop corrections below.

\subsection{Field-dependent masses}
\label{app:masses}

\paragraph{Neutral scalars.}
Because the portal term couples the two background directions, the 
CP-even neutral states mix. The mass matrix in the $(h,\phi)$ basis 
is
\begin{equation}
\mathcal{M}^2_S(h,\phi) =
\begin{pmatrix}
-\mu_H^2 + 3\lambda_H h^2 + \dfrac{\lambda_{H\phi}}{2}\phi^2 
& \lambda_{H\phi}\, h\,\phi \\[6pt]
\lambda_{H\phi}\, h\,\phi 
& \mu_\phi^2 + 3\lambda_\phi \phi^2 + \dfrac{\lambda_{H\phi}}{2}h^2
\end{pmatrix},
\label{eq:MS}
\end{equation}
whose eigenvalues $M_\pm^2(h,\phi)$ enter the loop sums with 
multiplicity $n_\pm = 1$ each. The off-diagonal entry vanishes in all 
three phases $\mathcal{P}_{\rm sym}$, $\mathcal{P}_{\rm int}$, and 
$\mathcal{P}_{\rm EW}$, but not along the tunnelling trajectories 
between them, where the full eigenvalues are used.

\paragraph{Goldstone bosons.}
The three Goldstone modes share
\begin{equation}
M_G^2(h,\phi) = -\mu_H^2 + \lambda_H h^2 
+ \frac{\lambda_{H\phi}}{2}\phi^2, 
\qquad n_G = 3.
\label{eq:MG}
\end{equation}
$M_G^2$ vanishes at the $T=0$ minimum, and the associated logarithmic 
sensitivity of $V_{\rm CW}$ is handled by evaluating the potential at 
$\mu = v$; the residual infrared effect on $T_c$ and $T_N$ is far 
below the thin-wall uncertainty already quoted for $\beta/H$.

\paragraph{Inert doublet.}
Both background fields feed the inert masses, the Higgs through the 
couplings of Eq.~\eqref{eq:Veta} and the singlet through 
$\lambda_{\phi\eta}$:
\begin{align}
M_{\eta^0_R}^2(h,\phi) &= \mu_\eta^2 
+ \frac{\lambda_3+\lambda_4+\lambda_5}{2}\,h^2 
+ \frac{\lambda_{\phi\eta}}{4}\,\phi^2, 
& n_{\eta^0_R} &= 1, \nonumber\\
M_{\eta^0_I}^2(h,\phi) &= \mu_\eta^2 
+ \frac{\lambda_3+\lambda_4-\lambda_5}{2}\,h^2 
+ \frac{\lambda_{\phi\eta}}{4}\,\phi^2, 
& n_{\eta^0_I} &= 1, \\
M_{\eta^\pm}^2(h,\phi) &= \mu_\eta^2 
+ \frac{\lambda_3}{2}\,h^2 
+ \frac{\lambda_{\phi\eta}}{4}\,\phi^2, 
& n_{\eta^\pm} &= 2. \nonumber
\label{eq:Metaapp}
\end{align}
At the benchmark value $\lambda_{\phi\eta} = 0.05$ the singlet 
contribution shifts the inert masses by less than $2~\mathrm{GeV}$ 
across the intermediate phase and is retained for completeness.

\paragraph{Gauge bosons and top quark.}
\begin{equation}
M_W^2(h) = \frac{g^2}{4}h^2, \qquad 
M_Z^2(h) = \frac{g^2+g'^2}{4}h^2, \qquad 
M_t^2(h) = \frac{y_t^2}{2}h^2,
\label{eq:Mgauge}
\end{equation}
with multiplicities $n_W = 6$, $n_Z = 3$, $n_\gamma = 2$, 
$n_t = -12$. The photon is massless at this order and contributes 
only through the longitudinal mixing described in 
Appendix~\ref{app:daisy}.

\subsection{One-loop zero-temperature and thermal pieces}
\label{app:oneloop}

With the spectrum above, the Coleman--Weinberg term is 
Eq.~\eqref{eq:VCW} in the $\overline{\rm MS}$ scheme at $\mu = v$, 
with $c_i = 3/2$ for scalars and fermions and $c_i = 5/6$ for gauge 
bosons. We do not impose on-shell counterterms; the one-loop shift of 
the $T=0$ vacuum is absorbed into the definition of the benchmark 
inputs. Along the bounce trajectory and near the origin, the field-dependent masses entering Eq.~\eqref{eq:VCW} can turn negative; we evaluate the Coleman--Weinberg term with the replacement 
$M_i^4 \ln(M_i^2/\mu^2) \to |M_i^2|^2\, \ln(|M_i^2|/\mu^2)$, retaining 
the real part, which is the standard prescription for the real 
effective potential and introduces no spurious imaginary 
contribution to the nucleation calculation. The thermal functions of Eq.~\eqref{eq:JBF} are evaluated 
numerically everywhere; for reference, their high-temperature 
expansions are
\begin{align}
J_B(x^2) &= -\frac{\pi^4}{45} + \frac{\pi^2}{12}x^2 
- \frac{\pi}{6}x^3 - \frac{x^4}{32}\ln\frac{x^2}{a_b} 
+ \mathcal{O}(x^6), 
& a_b &= 16\pi^2 e^{3/2 - 2\gamma_E}, \\
J_F(x^2) &= \frac{7\pi^4}{360} - \frac{\pi^2}{24}x^2 
- \frac{x^4}{32}\ln\frac{x^2}{a_f} + \mathcal{O}(x^6), 
& a_f &= \pi^2 e^{3/2 - 2\gamma_E},
\label{eq:JBFhighT}
\end{align}
the cubic term of $J_B$ being the origin of the thermally induced 
barriers of Section~\ref{subsec:twostep}.

\subsection{Thermal self-energies and daisy resummation}
\label{app:daisy}

The Arnold--Espinosa shift $M_i^2 \to M_i^2 + \Pi_i$ of 
Eq.~\eqref{eq:daisy} applies to all scalar degrees of freedom and to 
the \emph{longitudinal} gauge polarisations only; transverse modes 
receive no Debye mass at this order and are excluded from 
Eq.~\eqref{eq:Vdaisy}. The effective longitudinal multiplicities in 
the ring term are therefore $\bar{n}_{W_L} = 2$, $\bar{n}_{Z_L} = 
\bar{n}_{\gamma_L} = 1$, alongside the full scalar set.

The scalar self-energies, in the leading high-temperature 
approximation, are
\begin{align}
\Pi_H(T) &= T^2\left[\frac{3g^2}{16} + \frac{g'^2}{16} 
+ \frac{\lambda_H}{2} + \frac{y_t^2}{4} 
+ \frac{\lambda_{H\phi}}{24} 
+ \frac{2\lambda_3+\lambda_4}{12}\right],
\label{eq:PiH}\\
\Pi_\phi(T) &= T^2\left[\frac{\lambda_\phi}{4} 
+ \frac{\lambda_{H\phi}}{6} 
+ \frac{\lambda_{\phi\eta}}{12}\right],
\label{eq:Piphi}\\
\Pi_\eta(T) &= T^2\left[\frac{3g^2}{16} + \frac{g'^2}{16} 
+ \frac{\lambda_\eta}{2} 
+ \frac{2\lambda_3+\lambda_4}{12} 
+ \frac{\lambda_{\phi\eta}}{48}\right],
\label{eq:Pieta}
\end{align}
where $\Pi_H$ dresses both rows of Eq.~\eqref{eq:MS} associated with 
the doublet direction and the Goldstone mass 
Eq.~\eqref{eq:MG}, $\Pi_\phi$ the singlet entry, and $\Pi_\eta$ all 
three inert states. In practice the shift is applied to the diagonal 
of $\mathcal{M}^2_S$ before diagonalisation, so the resummed 
eigenvalues $\bar{M}_\pm^2$ feed the ring term.

For the gauge sector, with the SM field content plus one additional 
scalar doublet, the Debye masses are
\begin{equation}
\Pi_{W_L}(T) = 2\,g^2 T^2, \qquad 
\Pi_{B_L}(T) = 2\,g'^2 T^2,
\label{eq:PiWL}
\end{equation}
and the longitudinal $Z$--$\gamma$ system is diagonalised exactly:
\begin{equation}
\bar{M}^2_{Z_L/\gamma_L}(h,T) = 
\frac{1}{2}\left[\frac{g^2+g'^2}{4}h^2 + \Pi_{W_L} + \Pi_{B_L}\right] 
\pm \frac{1}{2}\sqrt{\left[\frac{g^2-g'^2}{4}h^2 
+ \Pi_{W_L} - \Pi_{B_L}\right]^2 + \frac{g^2 g'^2}{4}h^4},
\label{eq:ZgammaL}
\end{equation}
which reduces to $(M_Z^2, 0)$ at $T = 0$ and to the pure Debye 
masses at $h = 0$. The charged longitudinal mode carries 
$\bar{M}^2_{W_L} = M_W^2(h) + \Pi_{W_L}$.

The singlet, carrying no gauge charge, contributes to no Debye mass; 
its entire thermal footprint is the scalar loop content of 
Eqs.~\eqref{eq:PiH} and~\eqref{eq:Pieta}, and conversely the inert 
doublet stiffens the singlet direction through the 
$\lambda_{\phi\eta}/12$ term of Eq.~\eqref{eq:Piphi}. These 
cross-sector self-energies are the quantitative content of the 
statement in Section~\ref{subsec:veff} that the inert doublet 
participates in the resummation despite never acquiring a vacuum 
expectation value.

\section{Parameter space and benchmark selection}
\label{app:paramscan}

This appendix maps the viable parameter space of the singlet sector and 
justifies the benchmark choices in Table~\ref{tab:benchmarks}.

The two conditions that define the viable region are the singlet VEV 
condition $\mu_\phi^2 < 0$ (equivalently $\lambda_{H\phi} > 2m_\phi^2/v^2$) 
and the Espinosa--Quiros condition $\lambda_{H\phi} > \sqrt{2\lambda_H 
\lambda_\phi}$. For fixed $\lambda_\phi$, both conditions are lower bounds 
on $\lambda_{H\phi}$ as a function of $m_\phi$: the first rises as 
$m_\phi^2$, and the second is a constant horizontal line. The binding 
constraint switches from the Espinosa--Quiros condition at small $m_\phi$ 
to the $\mu_\phi^2 < 0$ condition at larger $m_\phi$.

\begin{figure}[h]
\centering
\includegraphics[width=0.80\textwidth]{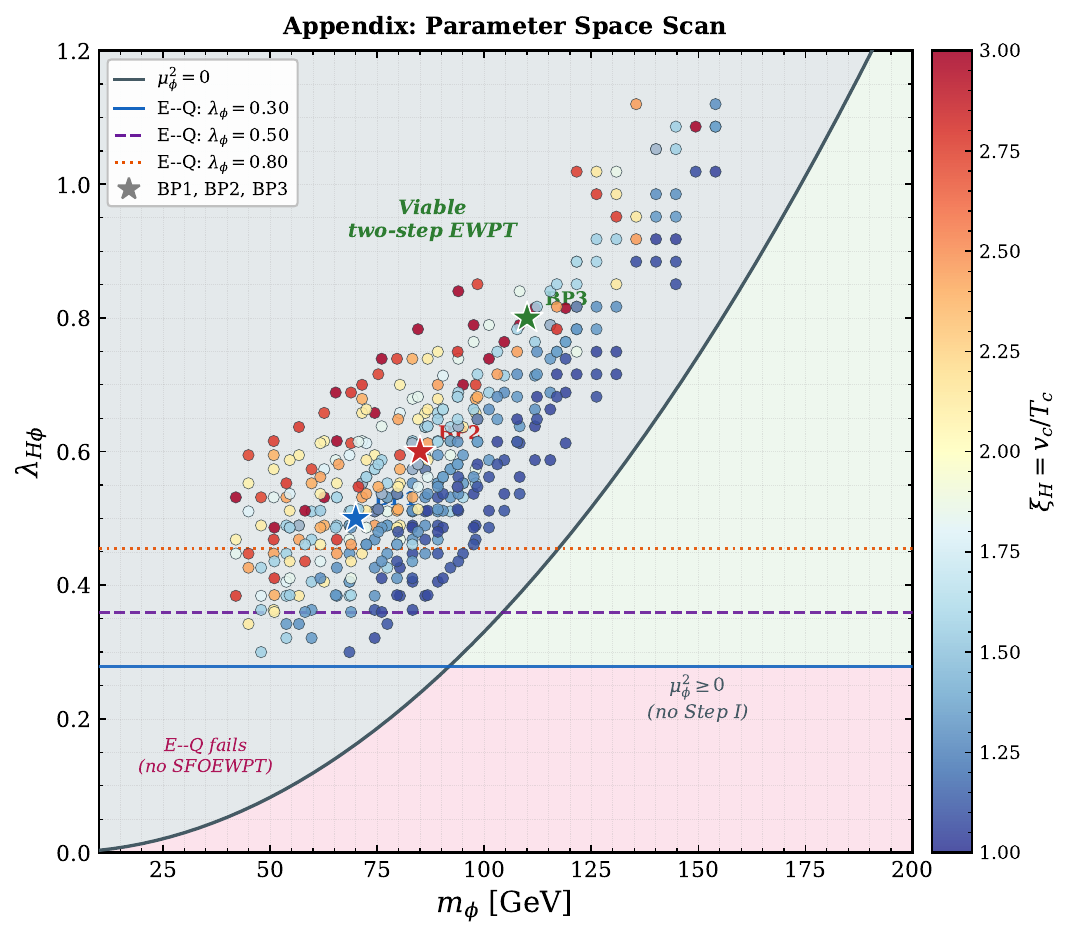}
\caption{Parameter space in the $(m_\phi, \lambda_{H\phi})$ plane. 
Grey region: $\mu_\phi^2 \geq 0$, Step~I of the phase transition is 
absent. Pink region: Espinosa--Quiros condition fails, Step~II is not 
strongly first-order. Green region: both conditions satisfied, two-step 
EWPT viable. Scan points from Section~\ref{sec:ewpt} are coloured by 
$\xi_H = v_c/T_c$. The three Espinosa--Quiros boundary lines correspond 
to $\lambda_\phi = 0.30$ (solid blue), $0.50$ (dashed purple), and $0.80$ 
(dotted orange). The grey curve marks $\mu_\phi^2 = 0$. Stars indicate 
the three benchmark points BP1, BP2, and BP3.}
\label{fig:paramscan}
\end{figure}

Figure~\ref{fig:paramscan} shows the resulting constraint map in the 
$(m_\phi, \lambda_{H\phi})$ plane. The grey region, bounded 
by the rising $\mu_\phi^2 = 0$ curve and occupying the lower-right of the 
plane, is excluded by the condition $\mu_\phi^2 \geq 0$: the singlet does 
not develop a vacuum expectation value at high temperature and Step~I of 
the phase transition is absent. The pink region, the 
horizontal strip below the lowest Espinosa--Quiros line, is excluded 
because that condition fails: even though $\mu_\phi^2 < 0$ may hold, the 
portal coupling is too weak to generate the barrier required for Step~II 
to be strongly first-order. The green region is viable for the two-step 
mechanism. Scan points from the full parameter scan of 
Section~\ref{sec:ewpt} are overlaid and coloured by $\xi_H = v_c/T_c$, 
showing that larger $\xi_H$ is achieved at larger $\lambda_{H\phi}$ and 
smaller $m_\phi$. Three Espinosa--Quiros boundary lines are shown for 
$\lambda_\phi = 0.30$, $0.50$, and $0.80$, illustrating how the viable 
region expands as $\lambda_\phi$ increases. The three benchmark points 
sample the low, intermediate, and high ends of the viable $m_\phi$ range.

Quantifying the selectivity of the two-step requirement 
across the sampled region: approximately $9\%$ of the parameter points 
evaluated in the scan complete the full two-step sequence with 
$\xi_H \geq 1$. Of the remainder, roughly $51\%$ fail because the singlet 
never develops an intermediate-phase vacuum (either $\mu_\phi^2 \geq 0$, 
so Step~I is absent, or the Step~I nucleation condition is not met), and 
roughly $40\%$ possess an intermediate phase but do not yield a genuine 
two-step history, either bypassing it in a single-step transition or 
generating too shallow a barrier for a strongly first-order Step~II. The 
two-step mechanism is therefore selective but not fine-tuned: the 
surviving points form the connected region shown rather than isolated 
islands. The scan samples the singlet parameters on dense grids in the 
$(\lambda_{H\phi}, m_\phi)$ plane, each centred on a benchmark and 
spanning $\pm 40\%$ about its central values, repeated across several 
values of $\lambda_\phi$. The $9\%$ survival fraction therefore refers to 
this benchmark-centred sampling and is representative of the neighbourhood 
of the viable region rather than of a uniform scan over the full ranges of 
Eq.~\eqref{eq:scangrid}; a blind uniform scan would return a smaller 
fraction, since much of that volume lies in the $\mu_\phi^2 \geq 0$ or 
Espinosa--Quiros-excluded regions.

For the IDM sector, the dependence on $\lambda_4$ and $\lambda_5$ enters 
only through the mass splittings $\Delta m$ and $\delta m$ defined in 
Eqs.~\eqref{eq:Deltam} and~\eqref{eq:deltam}. Since the relic density is 
governed by $\lambda_{H\eta}^{\rm eff} = \lambda_3 + \lambda_4 + \lambda_5$ 
and the direct detection cross section by the same combination, and since 
$|\lambda_4|, |\lambda_5| \ll \lambda_3 = 0.08$, the observables computed 
in this paper shift by less than $5\%$ across the range 
$\lambda_4 \in [0.01, 0.05]$ and $\lambda_5 \in [-0.10, -0.01]$ with 
$\lambda_5 < 0$.

\label{Bibliography}
\bibliographystyle{JHEP}
\bibliography{ref}

\end{document}